\documentclass[useAMS,usenatbib]{mn2e}

\usepackage{natbib}
\usepackage{graphicx}
\usepackage{amssymb}
\usepackage{color}
\usepackage[dvipsnames]{xcolor}
\usepackage{caption}
\usepackage{lipsum,graphicx,multicol}
\usepackage{float}
\usepackage[fleqn]{amsmath}
\usepackage{subcaption}

\title[AGN cool feedback and XRB analogy]{AGN cool feedback and analogy with X-ray binaries: \\ 
 from radiation pressure to cosmic ray driven outflows} 
\author[ ]
{W. Ishibashi$^{1}$\thanks{E-mail: wako.ishibashi@physik.uzh.ch} and A. C. Fabian$^{2}$
\footnotemark[0]\\
\footnotemark[0]\\
$^{1}$Physik-Institut, Universit$\ddot{a}$t Zurich, Winterthurerstrasse 190, 8057 Zurich, Switzerland \\
$^{2}$Institute of Astronomy, Madingley Road, Cambridge CB3 0HA 
}

\begin{document}

\pdfminorversion=4

\date{Accepted ? Received ?; in original form ? }

\pagerange{\pageref{firstpage}--\pageref{lastpage}} \pubyear{2012}

\maketitle

\label{firstpage}

\begin{abstract} 
Cool outflows are now commonly observed in galaxies, but their physical origin and driving mechanism remain unclear. Active galactic nucleus (AGN) feedback can potentially accelerate cool galactic outflows via cosmic rays (CR) and radiation pressure on dust. Here we  investigate the relative importance of CR and radiation feedback in AGNs, and we analyse the physical conditions for outflow launching as a function of the black hole accretion flow mode. We assume CRs from AGN jet origin and consider the analogy with Galactic X-ray binaries, whereby the jet is prominent at low accretion rates (hard state) and quenched at high accretion rates (soft state). We show that CR-driven outflows can be powered at low accretion rates and at large radii, whereas radiation pressure-driven outflows dominate at high accretion rates and small radii. Thus the two AGN feedback mechanisms --- CRs and radiation pressure on dust --- may play complementary roles in driving cool outflows on galactic scales. The transition from radiation pressure-driven outflows at higher accretion rates to CR-driven outflows at lower accretion rates likely corresponds to a transition in the underlying accretion flow modes (from a radiatively efficient accretion disc to a radiatively inefficient jet-dominated flow) over cosmic time. 
\end{abstract}  



\section{Introduction}
\label{Sect_Introduction}

There has been growing observational evidence of cool outflows ($T \lesssim 10^4$ K) on galactic scales, which can have important effects on the evolution of active galactic nuclei (AGN) and star-forming galaxies across cosmic time \citep[e.g. see the review by][and references therein]{Veilleux_et_2020}. The physical origin and driving mechanism of such cool galactic outflows --- including neutral, molecular, and dusty gas --- are still much debated. In principle, cool outflows can be directly driven by radiation pressure and cosmic ray pressure. Both radiation and cosmic rays can accelerate cold gas without the need of strong shocks, and thus naturally account for the existence of low-temperature material within galactic outflows. In this context, radiation pressure and cosmic rays may be considered as two `cool' feedback mechanisms.  

Radiation pressure on dust has been considered as a promising mechanism for AGN feedback \citep[][and references therein]{Fabian_1999, Fabian_2012}, and for driving outflows on galactic scales \citep[e.g.][]{Murray_et_2005, Thompson_et_2015}. In our previous studies, we have discussed how AGN `radiative dusty feedback' -- including radiation trapping -- can account for the dynamics and energetics observed in galactic molecular outflows \citep{Ishibashi_Fabian_2015, Ishibashi_et_2018a, Ishibashi_et_2021}. Radiation hydrodynamic (RHD) simulations have also been performed to follow the evolution of radiatively-driven winds in AGNs \citep{Bieri_et_2017, Costa_et_2018, Barnes_et_2020}. 

Cosmic rays (CR) provide another important source of feedback in galaxies. It has been realised since e.g. \citet{Ipavich_1975} that the CR pressure gradient can potentially accelerate cold gas in the form of galactic winds.  
CR-driven feedback has been gaining renewed interested in recent years
\citep[][]{Pfrommer_et_2017, Wiener_et_2017, Ehlert_et_2018, Hopkins_et_2020, Crocker_et_2021_b, Quataert_et_2022_a, Huang_Davis_2022, Heintz_Zweibel_2022, Thomas_et_2022}. The main astrophysical sites of CR production are supernova remnants in star-forming disc galaxies and relativistic jets in AGNs. 

CRs are high-energy charged particles and the CR energy density is dominated by $\gtrsim$GeV protons in galactic environments. Since CRs do not suffer much radiative loss, they can potentially drive cool gas out to large distances \citep{Wiener_et_2017}. The interaction between CRs and the ambient medium is mediated by magnetic fields. CRs scatter off small-scale magnetic fluctuations, which can be either self-excited by the CRs themselves (CR streaming) or produced by background turbulence (CR diffusion). The associated CR transport, i.e. how CRs propagate in galaxies and beyond, is currently a source of much debate in the field \citep[][and references therein]{Hopkins_et_2020, Quataert_et_2022_a}.  

The scattering process sets the CR mean free path $\lambda_\mathrm{CR} \sim c/\nu_\mathrm{CR}$, where $c$ is the speed of light and $\nu_\mathrm{CR}$ is the CR scattering rate. CRs typically have short mean free path (of the order of $\lambda_\mathrm{CR} \sim 1$ pc in the Galaxy), and their propagation may be modelled as a random walk \citep{Socrates_et_2008, Chan_et_2019}. The diffusion approximation may be particularly appropriate when CRs are scattered by ambient turbulence. The associated CR diffusion coefficient $\kappa_\mathrm{CR} \sim (1/3) c \lambda_\mathrm{CR}$ is uncertain and  difficult to constrain, but recent gamma-ray observations seem to favour rather large diffusivities, of the order of $\kappa_\mathrm{CR} \gtrsim 10^{29} \mathrm{cm^2 s^{-1}}$ \citep{Chan_et_2019}. 

CRs can be efficiently accelerated in AGN jets, whereby a fraction of the jet power is converted into CR luminosity \citep[e.g.][]{Sironi_Socrates_2010}. In the case of CR-dominated jets, the bulk of the jet energy can be injected in CR form \citep{Ruszkowski_et_2017}. The CR properties are then directly linked to the AGN jet properties, which in turn depend on the nature of the underlying accretion flow mode. 

In this context, the analogy with stellar-mass black holes in Galactic X-ray binaries (XRB) can give us important clues. Similarities in the radio jet and X-ray variability properties between XRB and AGN have been known for quite some time, leading to the so-called Fundamental Planes of black hole activity \citep{Merloni_et_2003, Falcke_et_2004, Koerding_et_2006, McHardy_et_2006}. These results suggest the scale-invariance of accreting black holes on all mass scales. 

Observations of spectral state transitions in XRBs indicate that a steady radio jet is present at low accretion rates (hard state), while the radio jet is quenched at high accretion rates (soft state) \citep[e.g.][and references therein]{Fender_2010}. It is well known that at high accretion rates, gas accretion proceeds via a standard geometrically thin and optically thick accretion disc with high radiative efficiency \citep{Shakura_Sunyaev_1973}. When the accretion rate falls below a certain critical value, there can be a transition to a radiatively inefficient accretion flow (RIAF) mode, such as advection-dominated accretion flow (ADAF) and variants thereof \citep[][and references therein]{Yuan_Narayan_2014}. By analogy with XRBs, two distinct accretion states may also occur in AGNs, with a strong jet present in the radiatively inefficient low-accretion regime and absent in the radiatively efficient high-accretion regime \citep{Churazov_et_2005}. 

AGN jet-mode feedback is commonly observed in the form of jet-inflated bubbles in clusters of galaxies \citep[][and references therein]{Fabian_2012}. The radio bubbles are detected as cavities in the X-ray surface brightness in cool core clusters. Bubbles rise buoyantly into the intracluster medium (ICM), subsequently evolving into ghost bubbles, as observed in the Perseus cluster. The bubble morphologies and sizes (with typical extent of $\sim 10$ kpc) vary depending on the interaction of the jet with the surrounding ambient medium \citep{McNamara_Nulsen_2012}. Although the exact bubble composition is still unknown, it is likely to involve a plasma of relativistic particles such as CR protons \citep{Dunn_Fabian_2004, Croston_et_2018}. The duty cycle of AGN jet bubbling is quite high, and the surrounding ICM can be heated by a more or less continuous bubbling process. A recent compilation of radio bubbles or X-ray cavities in different galaxy environments (clusters, groups, and ellipticals) is provided by \citet{Birzan_et_2020}. In some cases, scaled-down versions of bubbles may also be powered by stellar-mass black hole jets in XRB, as observed in Cygnus X-1 \citep{Gallo_et_2005, Russell_et_2007}. 

On the other hand, direct empirical evidence for AGN radiative-mode feedback is less clear-cut, due to the effects of obscuration. Nonetheless, there has been a huge observational progress in the detection of multi-phase galactic outflows in the past few years \citep{Fiore_et_2017, Fluetsch_et_2019, Veilleux_et_2020}. Galactic outflows can reach high velocities on kpc-scales, with the cold  molecular phase carrying the majority of the outflowing mass \citep{Fluetsch_et_2020}. Cool outflows are characterised by large momentum flux and high kinetic power, which require an efficient outflow driving mechanism. 

Here we study the relative importance of the two cool feedback mechanisms in AGNs -- cosmic rays vs. radiation pressure on dust -- as a function of the underlying accretion state. Based on the analogy with Galactic XRB, we analyse the accretion-dependent conditions for the launching of cool galactic outflows. 
The paper is structured as follows. We first introduce the Eddington limits for CR and radiation pressure on dust (Section \ref{Sect_Eddington_limits}), and then derive the physical conditions for outflow launching by CRs and radiation pressure, respectively (Section \ref{Sect_Outflow_conditions}). Assuming the analogy with XRBs, we investigate the accretion-dependent outflow driving mechanisms in Section \ref{Sect_XRB_analogy}, and compare the relative importance of CR feedback vs. radiation feedback in AGNs (Section \ref{Sect_CR-rad_feedback}). The coupling between CRs and dust grains in the circumgalactic medium is considered in Section \ref{Sect_CR-dust_CGM}. We further discuss the potential complementary (or supporting) roles of CR and radiation feedback, as well as the model limitations and future outlook (Section \ref{Sect_Discussion}), and conclude in Section \ref{Sect_Conclusion}. 


\section{Eddington limits}
\label{Sect_Eddington_limits}

The balance between the outward force due to CR or radiation pressure and the inward force due to gravity leads to the definition of a critical luminosity, which may be considered as a generalised form of the Eddington limit. The standard Eddington luminosity for electron scattering is given by $L_\mathrm{E} = (4 \pi G c m_p M_\mathrm{BH})/\sigma_T$, where $M_\mathrm{BH}$ is the black hole mass and $\sigma_T$ is the Thomson cross section. For radiation pressure on dust, the effective Eddington luminosity is much lower than the standard Eddington luminosity, because the dust absorption cross section is much larger than the Thomson cross section \citep[][]{Fabian_2012}. In the case of CR feedback, a CR Eddington limit can be defined as the critical CR flux above which hydrostatic equilibrium cannot be maintained \citep{Socrates_et_2008, Crocker_et_2021_b, Huang_Davis_2022, Heintz_Zweibel_2022}. Beyond this limit, CRs can potentially launch galactic outflows. Assuming that the energy released by the accretion process can be either delivered as radiation or injected as CRs (cf. Section \ref{Sect_XRB_analogy}), we derive the respective Eddington limits below. 


\subsection{Eddington limit for cosmic rays} 

The CR flux in the classical diffusion approximation is given by 
\begin{equation}
F_\mathrm{CR} = - \kappa_\mathrm{CR} \frac{d u_\mathrm{CR}}{dr}  , 
\end{equation}
where $u_\mathrm{CR} = 3 P_\mathrm{CR}$ is the CR energy density, $P_\mathrm{CR}$ is the CR pressure, and $\kappa_\mathrm{CR}$ is the CR diffusion coefficient. 
Balancing the CR pressure gradient and the gravitational force gives
\begin{equation}
\frac{dP_\mathrm{CR}}{dr} = - \frac{1}{3} \frac{F_\mathrm{CR}}{\kappa_\mathrm{CR}} = - \rho g  ,
\end{equation}
where $g = GM(r)/r^2$ is the gravitational acceleration and $M(r)$ is the enclosed mass. 
Thus the critical CR flux is given by
\begin{equation}
F_\mathrm{E,CR} = 3 \kappa_\mathrm{CR} \rho g .
\end{equation} 
Assuming that the gas follows an isothermal distribution with $\rho(r) = \frac{f_g \sigma^2}{2 \pi G r^2}$ (with gas fraction $f_g$ and velocity dispersion $\sigma$) and the standard flux-luminosity relation $F = L/4 \pi r^2$, we obtain the Eddington luminosity for CR
\begin{equation}
L_\mathrm{E,CR} = \frac{12 f_g \sigma^4}{G} \frac{\kappa_\mathrm{CR}}{r} .
\label{Eq_LE_CR}
\end{equation}
The numerical value of the CR Eddington luminosity for fiducial parameters is of the order of 
\begin{align}
L_\mathrm{E,CR} \cong 9.6 \times 10^{43} \; \mathrm{erg \, s^{-1}} 
\left( \frac{f_g}{0.1} \right) \left( \frac{\sigma}{200 \mathrm{km \, s^{-1}}} \right)^4 \nonumber \\
\times \left( \frac{\kappa_\mathrm{CR}}{10^{29} \mathrm{cm^2 \, s^{-1}}} \right) \left( \frac{r}{1 \mathrm{kpc}} \right)^{-1}  .
\end{align} 
We see that the CR Eddington luminosity decreases with increasing radius and decreasing CR diffusivity, scaling as $L_\mathrm{E,CR} \propto \kappa_\mathrm{CR}/r$. Therefore the CR Eddington limit can be more easily exceeded at larger radii and for smaller CR diffusion coefficients. 

\begin{figure}
\begin{center}
\includegraphics[angle=0,width=0.4\textwidth]{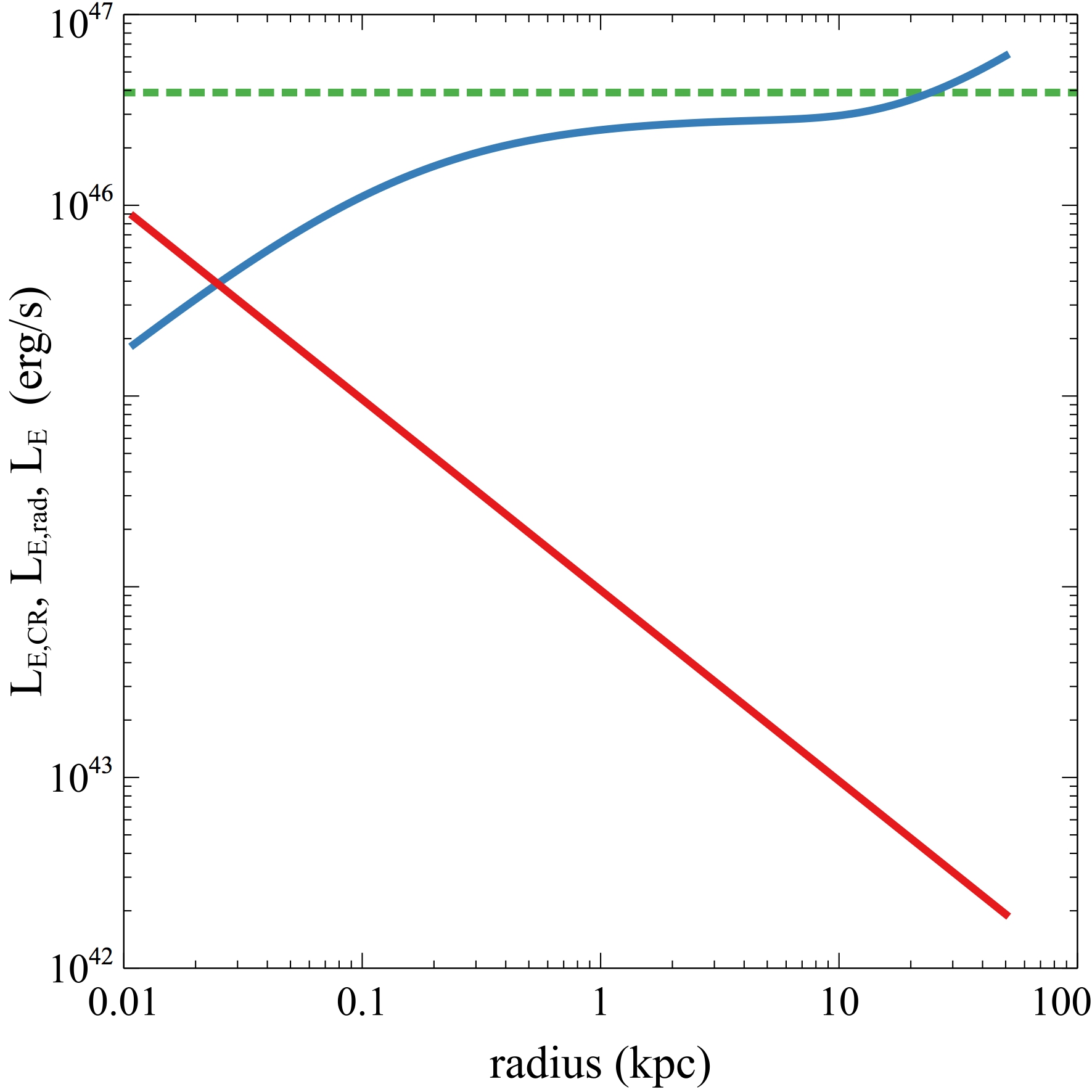} 
\caption{ Radial profiles of the different Eddington luminosities for an isothermal potential with velocity dispersion $\sigma = 200 \, \mathrm{km \, s^{-1}}$ and gas fraction $f_g = 0.1$. Standard Eddington luminosity for electron scattering (green dotted),  
CR Eddington luminosity with $\kappa_\mathrm{CR} = 10^{29} \mathrm{cm^{2} s^{-1}}$ (red line), radiative Eddington luminosity with $\kappa_\mathrm{IR}$=$5 \, \mathrm{cm^2 g^{-1}}$ and $\kappa_\mathrm{UV}$=$10^3 \, \mathrm{cm^2 g^{-1}}$ (blue curve). 
}
\label{Fig_LE_r}
\end{center}
\end{figure} 


\subsection{Eddington limit for radiation pressure on dust}

In the case of radiation feedback, a critical luminosity can be defined by equating the outward force due to radiation pressure and the inward force due to gravity. Assuming isothermal distribution, the radiative Eddington luminosity is given by \citep[e.g.][]{Ishibashi_Fabian_2015} 
\begin{equation}
L_\mathrm{E,rad} = \frac{4 f_g \sigma^4 c}{G} \left(1 + \tau_\mathrm{IR} - e^{-\tau_\mathrm{UV}} \right)^{-1} ,
\label{Eq_LE_rad}
\end{equation}
where
\begin{equation}
\tau_\mathrm{IR,UV} = \frac{\kappa_\mathrm{IR,UV} f_g \sigma^2}{2 \pi G r} ,
\end{equation}
are the infrared (IR) and ultraviolet (UV) optical depths, and $\kappa_\mathrm{IR}$=$5 \, \mathrm{cm^2 g^{-1} f_{dg, MW}}$ and $\kappa_\mathrm{UV}$=$10^3 \, \mathrm{cm^2 g^{-1} f_{dg, MW}}$ are the IR and UV opacities, with the dust-to-gas ratio normalised to the Milky Way value \citep{Ishibashi_Fabian_2015, Ishibashi_et_2018a}.  
Similar dust opacities are also commonly adopted in other works \citep{Murray_et_2011, Thompson_et_2015, Heckman_Thompson_2017}. 

Three distinct physical regimes can be identified depending on the optical depth of the surrounding medium: optically thick to both IR and UV at small radii, optically thick to UV but optically thin to IR (single scattering limit) at intermediate radii, and optically thin to UV at large radii. The optical depth falls off with increasing radius, with the IR and UV transparency radii given by $R_\mathrm{IR,UV} = \frac{\kappa_\mathrm{IR,UV} f_\mathrm{g} \sigma^2}{2 \pi G}$. We recall that the IR transparency radius is always smaller than the UV transparency radius ($R_\mathrm{IR} < R_\mathrm{UV}$). In the single scattering regime, the critical luminosity is a constant independent of radius ($L_\mathrm{E,rad}^\mathrm{SS} = \frac{4 f_g \sigma^4 c}{G}$); while the critical luminosity increases linearly with radius in the IR-optically-thick regime at small radii ($L_\mathrm{E,rad}^\mathrm{IR} = \frac{8 \pi c \sigma^2 r}{\kappa_\mathrm{IR}} \propto r$), and also in the UV-optically-thin regime at large radii ($L_\mathrm{E,rad}^\mathrm{UV} = \frac{8 \pi c \sigma^2 r}{\kappa_\mathrm{UV}} \propto r$). We note that in the IR radiation trapping regime, the effective Eddington limit is independent of the gas column density, and thus even dense material can be potentially disrupted \citep{Ishibashi_Fabian_2016b}.  


\subsection{Comparing CR and radiative Eddington limits}

The ratio between the Eddington luminosities for CR and radiation pressure is
\begin{equation}
\frac{L_\mathrm{E,CR}}{L_\mathrm{E,rad}} = \frac{3 \kappa_\mathrm{CR}}{c r}  \left(1 + \tau_\mathrm{IR} - e^{-\tau_\mathrm{UV}} \right) , 
\end{equation} 
which is about $L_\mathrm{E,CR}/L_\mathrm{E,rad} \cong  3.9 \times 10^{-3}$ for the following fiducial parameters: $\kappa_\mathrm{CR} = 10^{29} \mathrm{cm^{2} s^{-1}}$, $\sigma = 200 \, \mathrm{km \, s^{-1}}$, $f_g = 0.1$, $r = 1 \, \mathrm{kpc}$, $\kappa_\mathrm{IR}$=$5 \, \mathrm{cm^2 g^{-1}}$ and $\kappa_\mathrm{UV}$=$10^3 \, \mathrm{cm^2 g^{-1}}$. 

In Figure \ref{Fig_LE_r}, we show the CR Eddington luminosity (Eq. \ref{Eq_LE_CR}) and radiative Eddington luminosity (Eq. \ref{Eq_LE_rad}) as a function of radius. For comparison, the standard Eddington luminosity, assuming the $M_\mathrm{BH} - \sigma$ relation of \citet{Kormendy_Ho_2013}, is also shown (horizontal line). 
We see that $L_\mathrm{E,CR}$ decreases linearly with radius (as $\propto 1/r$), while $L_\mathrm{E,rad}$ tends to increase with radius following the different radial dependences in the three optical depth regimes. 
We observe that $L_\mathrm{E,CR}$ is (much) smaller than $L_\mathrm{E,rad}$ on galactic scales, and their difference increases with increasing radius. This suggests that CR-driven outflows can be more easily launched at larger radii compared to radiation pressure-driven outflows. 


\section{Conditions for outflow driving}
\label{Sect_Outflow_conditions}

In order to drive outflows on galactic scales, the CR and radiative luminosities must exceed their respective Eddington limits. This requires a minimum condition on the accretion rate (or equivalently a minimal Eddington ratio). 


\subsection{Radiation pressure-driven outflows}

Radiation from the central AGN is absorbed by dust grains, which transfer radiative momentum to the gas. Thus radiation pressure sweeps up the surrounding dusty gas into an outflowing shell. Radiation pressure acts on the bulk of the mass, and the radiative cooling of the dense dusty gas ensures that the outflow remains cool. 
Radiation pressure-driven outflows can develop if
\begin{equation}
L_\mathrm{rad} > L_\mathrm{E,rad} 
\label{Eq_rad_condition}
\end{equation}
where $L_\mathrm{rad} = \epsilon_\mathrm{rad} \dot{M} c^2$ is the radiative luminosity, with $\epsilon_\mathrm{rad}$ the radiative efficiency and $\dot{M}$ the mass accretion rate. 
Let us introduce the dimensionless accretion rate parameter $\dot{m} = \dot{M}/\dot{M}_E$, where $\dot{M}_E = L_E/(\eta c^2)$ is the Eddington accretion rate and $\eta$ is the accretion efficiency. 
The radiative luminosity is now expressed as $L_\mathrm{rad} = \frac{\epsilon_\mathrm{rad}}{\eta} \dot{m} L_E$. 
The condition for radiation pressure-driven outflows then requires a minimum accretion rate of
\begin{equation}
\dot{m} > \frac{\eta}{\epsilon_\mathrm{rad}} \frac{L_\mathrm{E,rad}}{L_\mathrm{E}} \equiv \dot{m}_\mathrm{rad} . 
\end{equation} 
This sets the minimal Eddington ratio ($\dot{m}_\mathrm{rad} \sim 0.64$ for fiducial parameters and with $\epsilon_\mathrm{rad} = \eta = 0.1$) required to launch radiation pressure-driven outflows on galactic scales. 


\subsection{CR-driven outflows}

Similarly, CR-driven outflows can develop provided that 
\begin{equation}
L_\mathrm{CR} > L_\mathrm{E,CR} 
\end{equation}
where $L_\mathrm{CR} = \epsilon_\mathrm{CR} L_\mathrm{jet}$ is the CR luminosity and $L_\mathrm{jet} = \epsilon_\mathrm{jet} \dot{M} c^2$ is the AGN jet luminosity. 
Here we consider CR from AGN jet origin, where a fraction of the jet power is injected as CR particles \citep[][]{Sironi_Socrates_2010}. 
Numerical simulations suggest that the jet mechanical efficiency is typically of the order a few percent, e.g. $\epsilon_\mathrm{jet} \sim 0.03$ \citep{Sadowski_Gaspari_2017}; while the CR conversion fraction may be of the order of $\epsilon_\mathrm{CR} \sim 0.8$ for a CR-dominated jet \citep{Ruszkowski_et_2017}. 
The condition for CR-driven outflows then leads to 
\begin{equation}
\dot{m} > \frac{\eta}{\epsilon_\mathrm{CR} \epsilon_\mathrm{jet}} \frac{L_\mathrm{E,CR}}{L_\mathrm{E}} \equiv \dot{m}_\mathrm{CR} . 
\end{equation}
We note that the minimal Eddington ratio for launching CR-driven outflows ($\dot{m}_\mathrm{CR} \sim 0.01$ for fiducial parameters) is smaller than the corresponding value for radiation pressure. 
The evolution of CR-driven shells may by analogous to that of interstellar wind bubbles \citep[e.g.][]{Weaver_et_1977}, but with the driving energy directly tapped from the central AGN. 


\subsection{Comparing $\dot{m}_\mathrm{CR}$ and $\dot{m}_\mathrm{rad}$}

\begin{figure}
\begin{center}
\includegraphics[angle=0,width=0.4\textwidth]{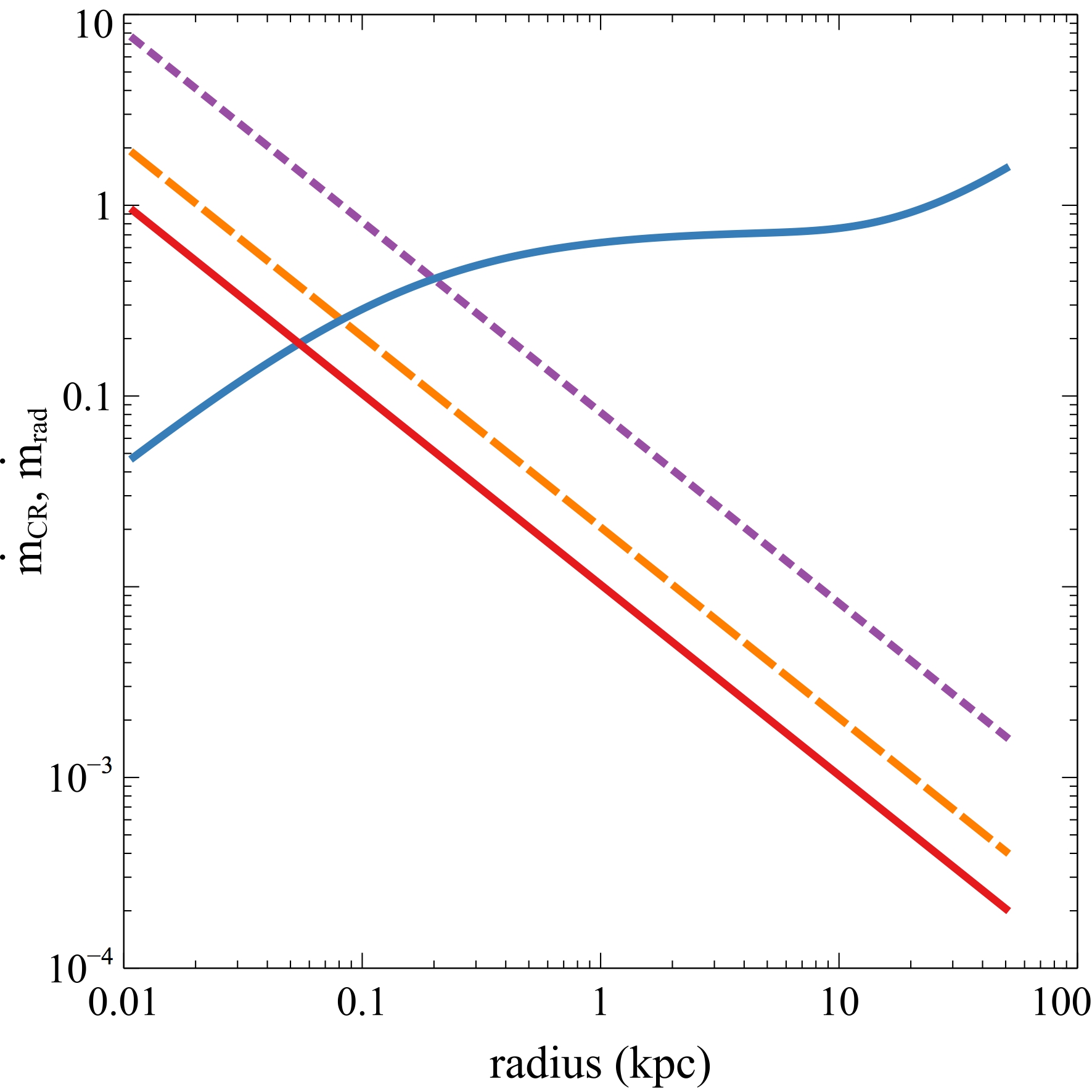} 
\caption{ 
Minimum accretion rates (or minimal Eddington ratios) for outflow driving by CRs ($\dot{m}_\mathrm{CR}$, coloured lines) and radiation pressure ($\dot{m}_\mathrm{rad}$, blue curve) as a function of radius. 
Variations in the CR injection fraction: $\dot{m}_\mathrm{CR}$ with $\epsilon_\mathrm{CR} = 0.8$ (red solid),  $\epsilon_\mathrm{CR} = 0.4$ (orange dashed), $\epsilon_\mathrm{CR} = 0.1$ (violet dotted).  
}
\label{Fig_min-mdot_r}
\end{center}
\end{figure} 

In Figure \ref{Fig_min-mdot_r}, we plot the minimal Eddington ratios required for outflow driving by CRs and radiation pressure, as a function of radius. We see that radiation-driven outflows require relatively high accretion rates (at significant fractions of the Eddington rate), whereas CR outflows can be powered by much lower accretion rates at large radii. 
In fact, the minimal Eddington ratio for CR is always smaller than that for radiation pressure ($\dot{m}_\mathrm{CR} < \dot{m}_\mathrm{rad}$) on galactic scales, and their difference increases with increasing radius. This is because $\dot{m}_\mathrm{CR}(r)$ decreases linearly with increasing radius ($\propto 1/r$); whereas $\dot{m}_\mathrm{rad}(r) \propto L_\mathrm{E,rad}(r)$ grows with radius (as $\propto r$) in the IR-optically-thick and UV-optically-thin regimes, while it is independent of radius in the single scattering limit (see the different radial profiles of the radiative Eddington luminosity in the three optical depth regimes described in Section \ref{Sect_Eddington_limits}). We also observe that $\dot{m}_\mathrm{CR}$ is smaller for larger CR conversion fraction ($\epsilon_\mathrm{CR}$), such that the crossing with $\dot{m}_\mathrm{rad}$ occurs at a smaller radius.  

The ratio between the two minimal Eddington ratios is given by
\begin{equation}
\frac{\dot{m}_\mathrm{CR}}{\dot{m}_\mathrm{rad}} = \frac{\epsilon_\mathrm{rad}}{\epsilon_\mathrm{CR} \epsilon_\mathrm{jet}} \frac{L_\mathrm{E,CR}}{L_\mathrm{E,rad}} , 
\end{equation}
which is always smaller than unity on galactic scales (e.g. $\dot{m}_\mathrm{CR}/\dot{m}_\mathrm{rad} \sim 0.016$ at $r \sim 1$ kpc).
Indeed, the minimum accretion rate required for outflow driving is much lower for CRs than for radiation pressure on dust -- except at the innermost radii. 
As a consequence, CR-driven outflows can be more easily launched by lower accretion rates at larger radii, and vice-versa.   


\section{Analogy with Galactic X-ray binaries: accretion-dependence}
\label{Sect_XRB_analogy}

The total accretion power ($L_\mathrm{tot} = \eta \dot{M} c^2$) released by the accreting black hole is ultimately distributed between two feedback modes. The relative importance of the radiative feedback vs. kinetic feedback depends on the accretion rate \citep{Churazov_et_2005, Merloni_Heinz_2008, Mocz_et_2013}. 
In the case of Galactic XRBs, there is a clear correlation between the radio jet power and the X-ray spectral state. At low accretion rates (hard state), a major fraction of the accretion power is released in kinetic form and a steady radio jet is present; whereas at high accretion rates (soft state), the radiative output dominates and the radio jet is quenched \citep{Fender_et_2004, Fender_Munoz_2016}. 

By analogy with XRBs, we assume that the relative importance of the AGN radiative and kinetic outputs also depends on the accretion rate. Considering that the total power is split between radiative and kinetic forms, the power balance gives: $L_\mathrm{tot} = L_\mathrm{rad} + L_\mathrm{jet}$, and hence $\eta = \epsilon_\mathrm{rad}(\dot{m}) + \epsilon_\mathrm{jet}(\dot{m})$, where the radiative and kinetic efficiencies are now both functions of the accretion rate. At low accretion rates, much of the accretion power goes into the jet outflow and the radiative efficiency is low in such a hot RIAF-like flow. Above a certain critical accretion rate ($\dot{m}_c$), corresponding to the transition to a standard accretion disc, the jet power drops and the radiative power becomes dominant \citep{Churazov_et_2005}. More specifically, the radiative efficiency is a constant (equal to the accretion efficiency) in the radiatively efficient accretion disc regime; while below the critical accretion rate, the radiative efficiency decreases with decreasing accretion rate as \citep[e.g.][]{Merloni_Heinz_2008}
\begin{equation}
\epsilon_\mathrm{rad} (\dot{m}) = \eta \hspace{2.5cm}  \mathrm{for} \; \dot{m} \geq \dot{m}_c 
\end{equation}
\begin{equation}
\epsilon_\mathrm{rad} (\dot{m}) = \eta \left( \frac{\dot{m}}{\dot{m}_c} \right) \hspace{1.4cm} \mathrm{for} \; \dot{m} < \dot{m}_c 
\end{equation} 
where $\dot{m}_c$ is the critical accretion rate defining the transition between the radiatively efficient accretion disc and the radiatively inefficient accretion flow. As a consequence, the jet efficiency $\epsilon_\mathrm{jet}(\dot{m}) = \eta - \epsilon_\mathrm{rad}(\dot{m})$ in the two accretion regimes is given by:
\begin{equation}
\epsilon_\mathrm{jet} (\dot{m}) = 0 \hspace{3cm} \mathrm{for} \; \dot{m} \geq \dot{m}_c 
\end{equation}
\begin{equation}
\epsilon_\mathrm{jet} (\dot{m}) = \eta \left( 1- \frac{\dot{m}}{\dot{m}_c} \right) \hspace{1.3cm} \mathrm{for} \; \dot{m} < \dot{m}_c 
\end{equation}
A similar split between radiative and kinetic forms as a function of the accretion rate is also employed in numerical simulations \citep{Sadowski_Gaspari_2017, Qiu_et_2019}. 

\begin{figure}
\begin{center}
\includegraphics[angle=0,width=0.4\textwidth]{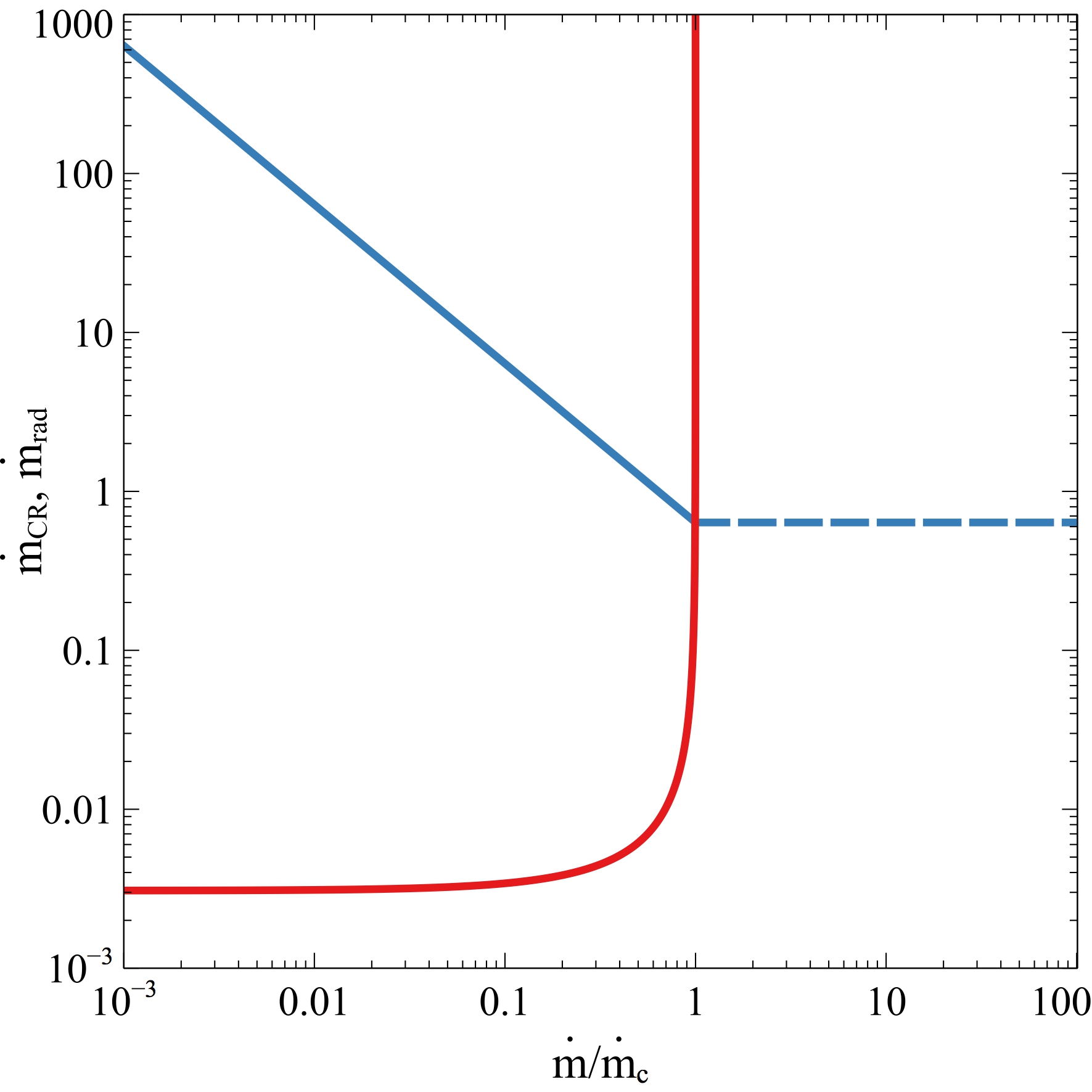} 
\caption{ 
Minimal Eddington ratios for outflow driving by CRs ($\dot{m}_\mathrm{CR}$, red) and radiation pressure ($\dot{m}_\mathrm{rad}$ for $\dot{m} < \dot{m}_c$, blue solid; and $\dot{m}_\mathrm{rad}$ for $\dot{m} > \dot{m}_c$, blue dashed). }
\label{Fig_min-mdot_acc}
\end{center}
\end{figure} 


\subsection{Outflow driving with accretion-dependence}

The development of galactic outflows now explicitly depends on the accretion rate, and the outflow launching conditions should be adapted accordingly. Including the accretion-dependence, the condition for radiation pressure-driven outflows is given by 
\begin{equation}
\dot{m} > \dot{m}_\mathrm{rad} = \frac{L_\mathrm{E,rad}}{L_\mathrm{E}}  \hspace{2cm} \mathrm{for} \; \dot{m} \geq \dot{m}_c
\end{equation}
\begin{equation}
\dot{m} > \dot{m}_\mathrm{rad} = \frac{1}{\dot{m}/\dot{m}_c} \frac{L_\mathrm{E,rad}}{L_\mathrm{E}} \hspace{1.05cm} \mathrm{for} \; \dot{m} < \dot{m}_c
\end{equation}
Similarly, CR-driven outflows can be launched in the low-accretion regime provided that 
\begin{equation}
\dot{m} > \dot{m}_\mathrm{CR} = 
\frac{1}{\epsilon_\mathrm{CR}(1-\dot{m}/\dot{m}_c)} \frac{L_\mathrm{E,CR}}{L_\mathrm{E}} \hspace{1cm} \mathrm{for} \; \dot{m} < \dot{m}_c
\end{equation}

In Figure \ref{Fig_min-mdot_acc}, we show the minimal Eddington ratios for outflow driving as a function of the accretion fraction ($\dot{m}/\dot{m}_c$). We see that $\dot{m}_\mathrm{CR}$ is small in the low-accretion regime, but increases steeply for $\dot{m}/\dot{m}_c \rightarrow 1$. In contrast, $\dot{m}_\mathrm{rad}$ is very large (super-Eddington) at low accretion rates, but decreases with increasing accretion rate and becomes a constant in the high-accretion regime. Comparing the two, we note that $\dot{m}_\mathrm{CR}$ is always smaller than $\dot{m}_\mathrm{rad}$ (for $\dot{m}/\dot{m}_c < 1$) in the radiatively inefficient jet-dominated regime as expected. Therefore in the low-accretion regime, CR-driven outflows can be more easily launched than radiation-driven outflows; whereas the situation is reversed at high accretion rates. 

\begin{figure*}
\begin{multicols}{2}
    \includegraphics[width=0.8\linewidth]{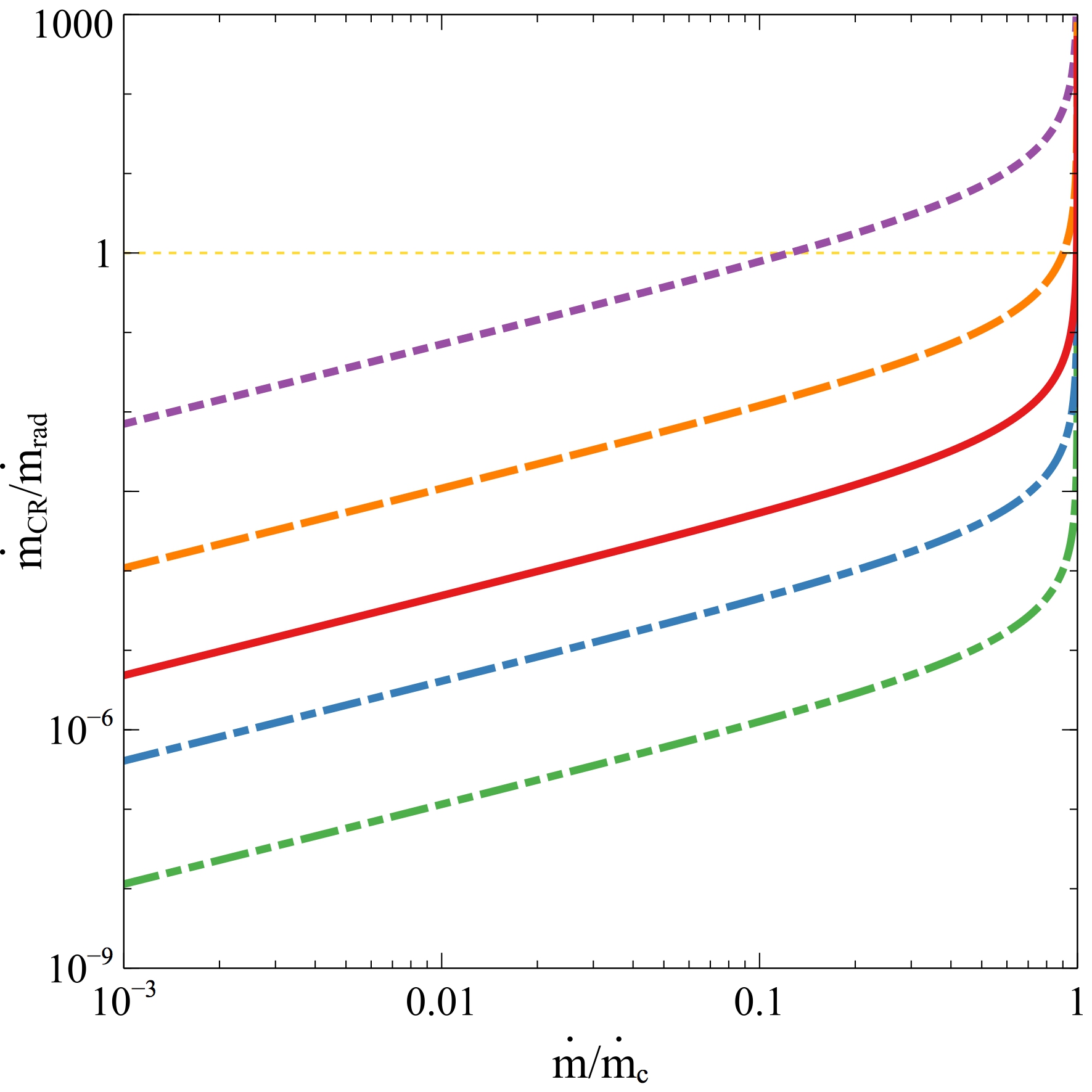}\par
    \includegraphics[width=0.8\linewidth]{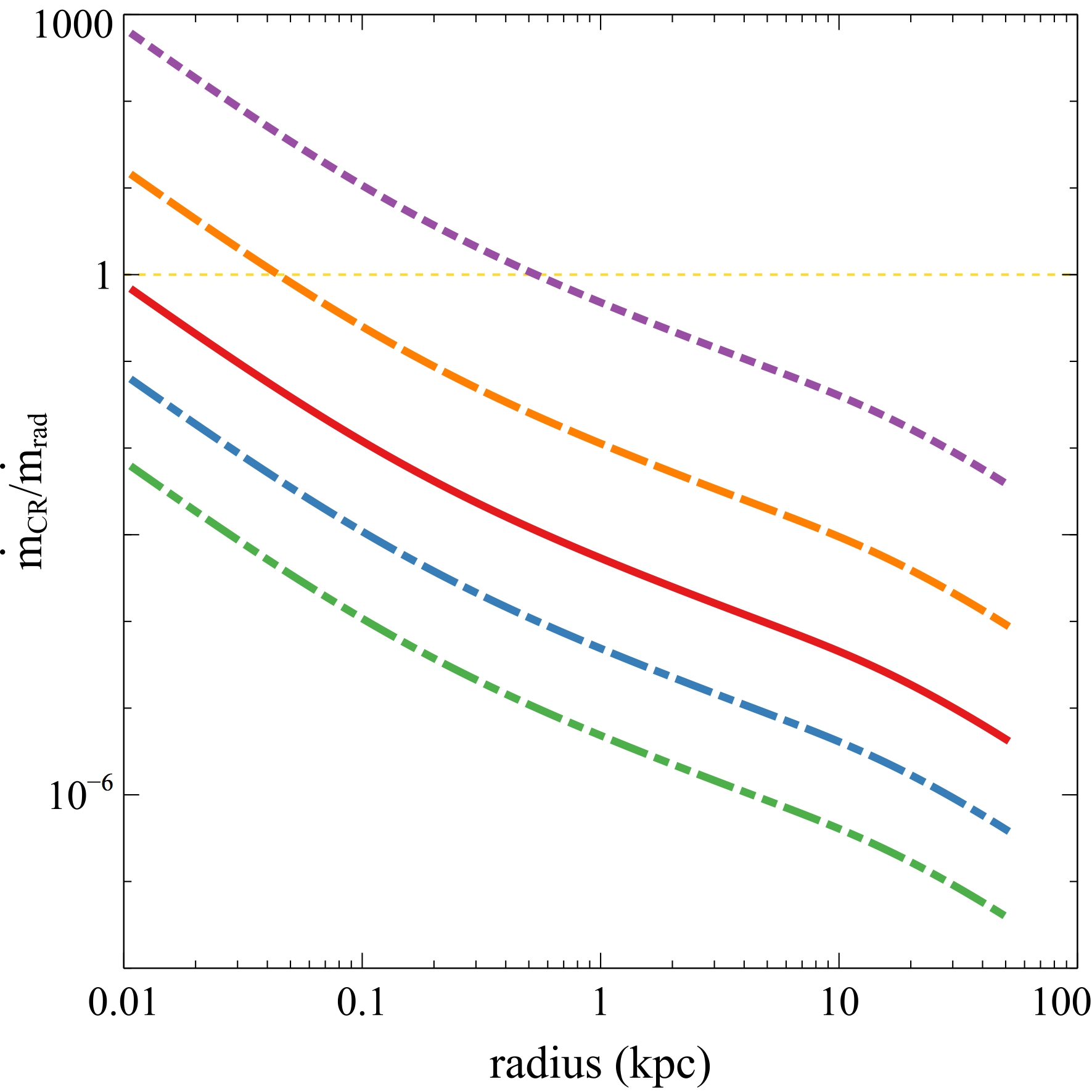}\par 
    \end{multicols}
\caption{
The $\dot{m}_\mathrm{CR}/\dot{m}_\mathrm{rad}$ ratio as a function of the accretion fraction $\dot{m}/\dot{m}_c$ (left-hand panel) and radius $r$ (right-hand panel). 
Variations in radius (left-hand panel):  $r = 0.01$ kpc (violet dotted), $r = 0.1$ kpc (orange dashed), $r = 1$ kpc (red solid), $r = 10$ kpc (blue dash-dot), $r = 100$ kpc (green dash-dot-dot). 
Variations in accretion fraction (right-hand panel): $\dot{m}/\dot{m}_c = 10^{-3}$ (green dash-dot-dot),  $\dot{m}/\dot{m}_c = 10^{-2}$ (blue dash-dot), $\dot{m}/\dot{m}_c = 0.1$ (red solid), $\dot{m}/\dot{m}_c = 0.7$ (orange dashed), $\dot{m}/\dot{m}_c = 0.9$ (violet dotted).  
The horizontal yellow lines mark the limit $\dot{m}_\mathrm{CR}/\dot{m}_\mathrm{rad} = 1$ in both panels. } 
\label{Fig_mdotratio_acc_r}
\end{figure*}

The ratio between the minimal CR and radiative Eddington ratios for outflow driving is 
\begin{equation}
\frac{\dot{m}_\mathrm{CR}}{\dot{m}_\mathrm{rad}} = \frac{1}{\epsilon_\mathrm{CR}} \frac{(\dot{m}/\dot{m}_c)}{1 - (\dot{m}/\dot{m}_c)} \frac{L_\mathrm{E,CR}}{L_\mathrm{E,rad}}
\end{equation}
In Figure \ref{Fig_mdotratio_acc_r}, we show the $\dot{m}_\mathrm{CR}/\dot{m}_\mathrm{rad}$ ratio as a function of the accretion fraction $\dot{m}/\dot{m}_c$ (left-hand panel) and as a function of radius $r$ (right-hand panel).   
We see that the $\dot{m}_\mathrm{CR}/\dot{m}_\mathrm{rad}$ ratio is mostly lower than unity, except at the smallest radii and highest $\dot{m}/\dot{m}_c$. The ratio decreases with increasing radius and decreasing $\dot{m}/\dot{m}_c$. By combining the two trends, we observe that it is easier to launch CR-driven outflows at low accretion rates and at larger radii. Conversely, radiation pressure-driven outflows dominate at high accretion rates and smaller radii. 

\section{Cosmic ray vs. radiation feedback }
\label{Sect_CR-rad_feedback}

We next compare the relative importance of CRs and radiation pressure as feedback mechanisms in galaxies, and their complementary roles in powering cool galactic outflows.


\subsection{CR and radiative luminosities }

The CR and radiative luminosities with $\dot{m}$-dependence can be computed in the respective accretion regimes. 
The accretion-dependent radiative luminosity is given by
\begin{equation}
L_\mathrm{rad}(\dot{m}) = \dot{m} L_E   \hspace{2cm} \mathrm{for} \; \dot{m} \geq \dot{m}_c
\end{equation} 
\begin{equation}
L_\mathrm{rad}(\dot{m}) = \left( \frac{\dot{m}}{\dot{m}_c} \right) \dot{m} L_E  \hspace{1cm} \mathrm{for} \; \dot{m} < \dot{m}_c
\end{equation}   
Likewise, the CR luminosity in the low-accretion regime is 
\begin{equation}
L_\mathrm{CR}(\dot{m}) = \epsilon_\mathrm{CR} \left( 1-\frac{\dot{m}}{\dot{m}_c} \right) \dot{m} L_E \hspace{1cm} \mathrm{for} \; \dot{m} < \dot{m}_c
\end{equation}
Observations and theoretical considerations suggest that the value of the critical accretion rate typically lies between $\sim 10^{-3}$ and $\sim 10^{-1}$ \citep{Churazov_et_2005, Koerding_et_2008, Merloni_Heinz_2008, Mocz_et_2013}, and here we take $\dot{m}_c = 0.01$ as a fiducial value.

\begin{figure}
\begin{center}
\includegraphics[angle=0,width=0.4\textwidth]{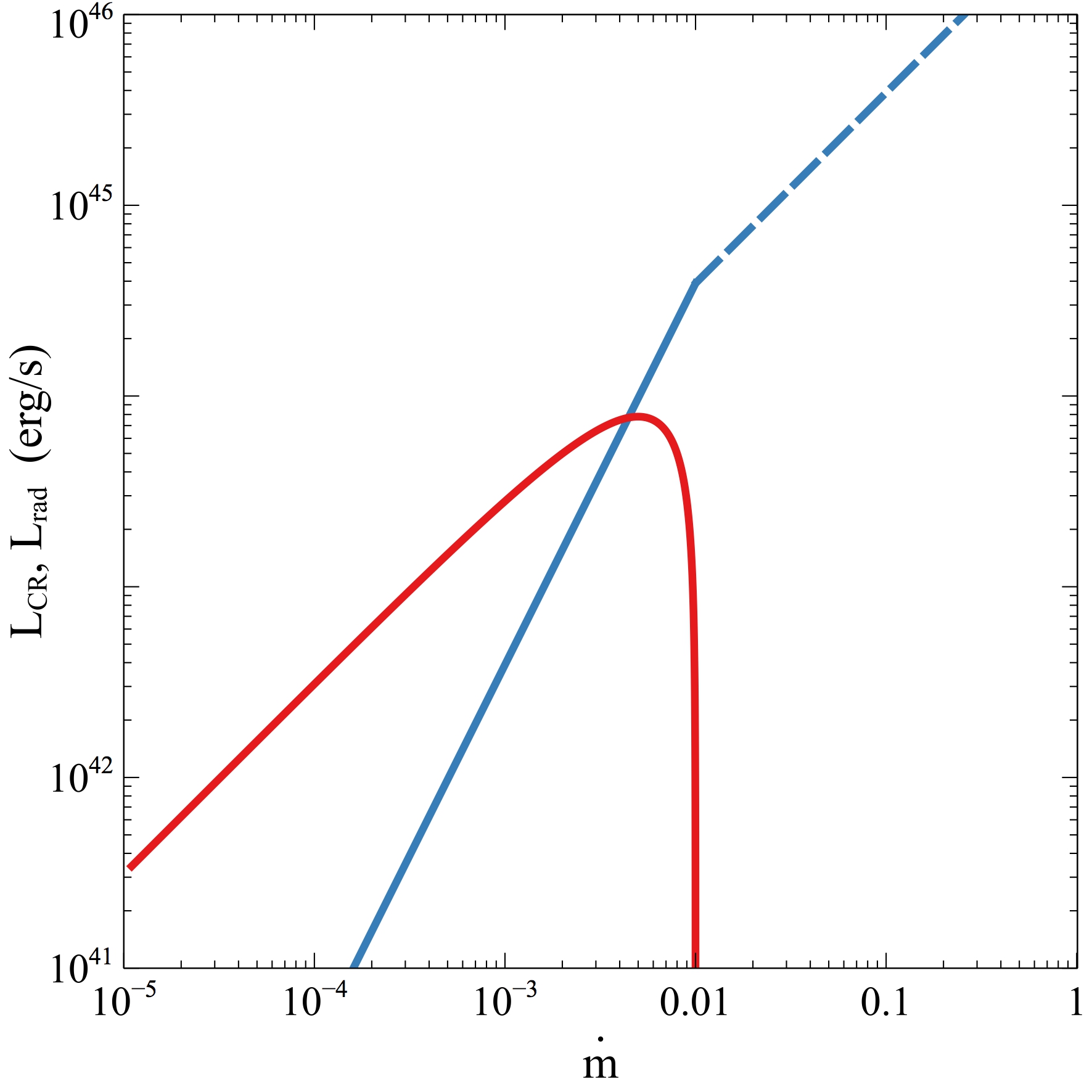} 
\caption{ CR luminosity ($L_\mathrm{CR}$, red solid) and radiative luminosity ($L_\mathrm{rad}$ for $\dot{m} < \dot{m}_c$, blue solid; and $L_\mathrm{rad}$ for $\dot{m} > \dot{m}_c$, blue dashed) as a function of the accretion rate $\dot{m}$. 
Note that the CR luminosity scales with accretion rate as $L_\mathrm{CR}(\dot{m}) \propto \dot{m} - \dot{m}^2$; while the radiative luminosity increases quadratically as $L_\mathrm{rad}(\dot{m}) \propto \dot{m}^2$ for $\dot{m} < \dot{m}_c$, and then grows linearly as $L_\mathrm{rad}(\dot{m}) \propto \dot{m}$ for $\dot{m} \geq \dot{m}_c$.
}
\label{Fig_LCR_Lrad_m}
\end{center}
\end{figure}

In Figure \ref{Fig_LCR_Lrad_m}, we show the resulting CR and radiative luminosities as a function of the accretion rate. We see that the CR luminosity is greater than the radiative luminosity at low accretion rates ($\dot{m} < \dot{m}_c$), but $L_\mathrm{CR}$ steeply falls off for $\dot{m} \rightarrow \dot{m}_c$. The radiative luminosity exceeds the CR luminosity for $\dot{m} > \dot{m}_c$, and takes over at high accretion rates. In fact, $L_\mathrm{rad}$ becomes the only dominant component in the high-accretion regime. The crossing between the two feedback modes occurs around the critical accretion rate.

We can also compute the ratio between the CR and radiative luminosities
\begin{equation}
\frac{L_\mathrm{CR}}{L_\mathrm{rad}} = \epsilon_\mathrm{CR} \frac{1-\dot{m}/\dot{m_c}}{\dot{m}/\dot{m_c}} \hspace{1cm} \mathrm{for} \; \dot{m} < \dot{m}_c
\end{equation}
From Figure \ref{Fig_L_P_m} (left-hand panel), we see that the $L_\mathrm{CR}/L_\mathrm{rad}$ ratio is greater than unity for $\dot{m} <  \dot{m}_c$, and is larger for larger critical accretion rate $\dot{m}_c$. 
Thus the CR luminosity dominates the radiative luminosity in the low-accretion state, while the radiative output becomes the sole dominant mode in the high-accretion regime. 


\subsection{CR and radiation pressures}

The CR and radiation pressures acting on the gas can also be directly computed. 
The radiation pressure is defined as
\begin{equation}
P_\mathrm{rad} = \frac{L_\mathrm{rad}}{4 \pi r^2 c} , 
\end{equation} 
while the CR pressure is given by
\begin{equation}
P_\mathrm{CR} = \frac{L_\mathrm{CR}}{4 \pi r^2 v_\mathrm{CR}} . 
\end{equation}
The characteristic velocity $v_\mathrm{CR}$ in the diffusion limit is  
\begin{equation}
v_\mathrm{CR} \sim \frac{c}{\tau_\mathrm{CR}} \sim \frac{\kappa_\mathrm{CR}}{H} , 
\end{equation} 
where $\tau_\mathrm{CR}$ is the CR optical depth and $H$ ($\sim 1$ kpc) is a characteristic scale height \citep{Socrates_et_2008}. 
Thus the CR pressure can be written as
\begin{equation}
P_\mathrm{CR} = \frac{H}{\kappa_\mathrm{CR}} \frac{L_\mathrm{CR}}{4 \pi r^2 } . 
\end{equation}
Taking into account the accretion-dependence, the CR pressure in the low-accretion regime is given by
\begin{equation}
P_\mathrm{CR}(\dot{m}) =  \frac{H}{\kappa_\mathrm{CR}} \frac{1}{4 \pi r^2 }  \epsilon_\mathrm{CR} \left( 1-\frac{\dot{m}}{\dot{m}_c} \right) \dot{m} L_E
\hspace{0.8cm} \mathrm{for} \; \dot{m} < \dot{m}_c
\end{equation}  
which scales as $\propto (\dot{m} - \dot{m}^2)$, while $P_\mathrm{CR}(\dot{m}) = 0$ for $\dot{m} \geq \dot{m}_c$. 
The radiation pressure in the high-accretion regime is 
\begin{equation}
P_\mathrm{rad}(\dot{m}) = \frac{\dot{m} L_E}{4 \pi r^2 c} 
\hspace{2cm} \mathrm{for} \; \dot{m} \geq \dot{m}_c
\end{equation} 
which scales linearly with the accretion rate ($\propto \dot{m}$); whereas the radiation pressure in the low-accretion regime is given by 
\begin{equation}
P_\mathrm{rad}(\dot{m}) = \frac{(\dot{m}/\dot{m}_c) \dot{m} L_E}{4 \pi r^2 c} 
\hspace{1cm} \mathrm{for} \; \dot{m} < \dot{m}_c
\end{equation} 
with a quadratic dependence on the accretion rate ($\propto \dot{m}^2$). 

\begin{figure*}
\begin{multicols}{2}
    \includegraphics[width=0.8\linewidth]{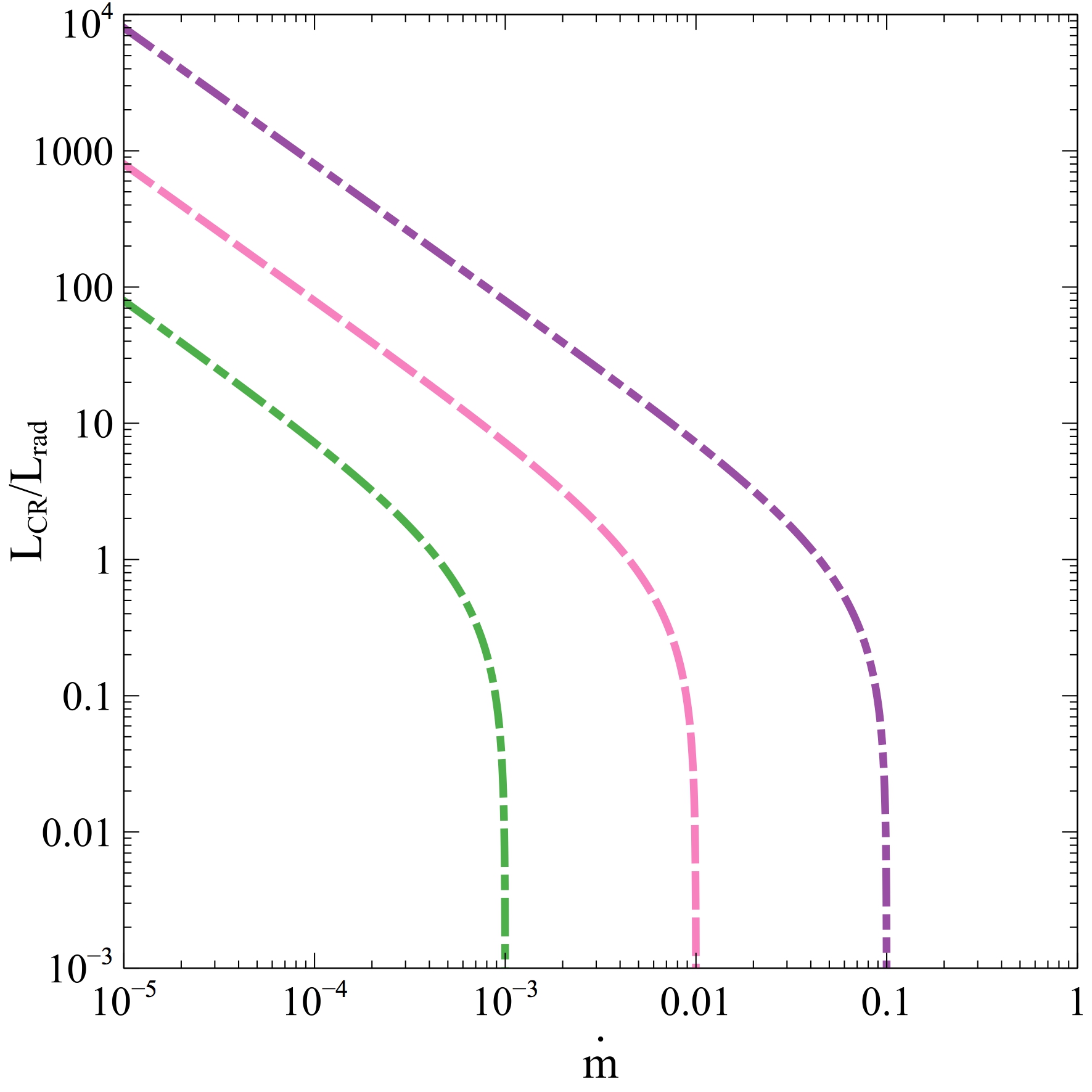}\par
    \includegraphics[width=0.8\linewidth]{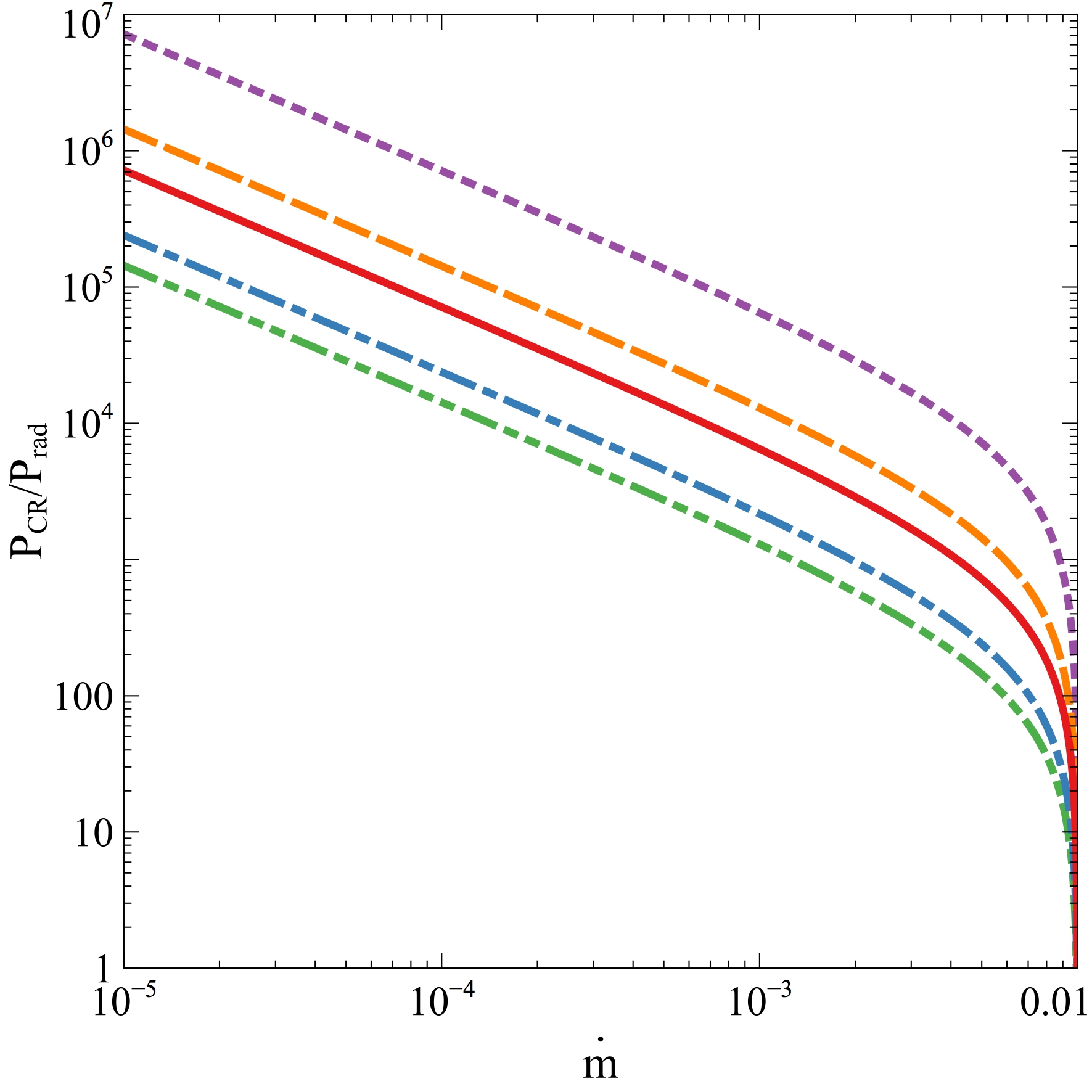}\par 
    \end{multicols}
\caption{ 
Left-hand panel: the CR-to-radiative luminosity ratio $L_\mathrm{CR}/L_\mathrm{rad}$ as a function of the accretion rate $\dot{m}$,  with variations in the critical accretion rate: $\dot{m}_c = 10^{-3}$ (green dash-dot), $\dot{m}_c = 0.01$ (pink dashed), $\dot{m}_c = 0.1$ (violet dash-dot-dot).
Right-hand panel: the ratio of CR and radiation pressures $P_\mathrm{CR}/P_\mathrm{rad}$ as a function of the accretion rate $\dot{m}$, with variations in the CR diffusion coefficient: $\kappa_\mathrm{CR} = 10^{28} \mathrm{cm^2/s}$ (violet dotted), $\kappa_\mathrm{CR} = 5 \times 10^{28} \mathrm{cm^2/s}$ (orange dashed), $\kappa_\mathrm{CR} = 10^{29} \mathrm{cm^2/s}$ (red solid), $\kappa_\mathrm{CR} = 3 \times 10^{29} \mathrm{cm^2/s}$ (blue dash-dot), $\kappa_\mathrm{CR} = 5 \times 10^{29} \mathrm{cm^2/s}$ (green dash-dot-dot).
} 
\label{Fig_L_P_m}
\end{figure*}

The ratio between the CR and radiation pressures in the low-accretion regime is given by
\begin{equation}
\frac{P_\mathrm{CR}}{P_\mathrm{rad}} = \frac{H c }{\kappa_\mathrm{CR}} \epsilon_\mathrm{CR}  \frac{1-\dot{m}/\dot{m}_c}{\dot{m}/\dot{m}_c}
\end{equation} 
Figure \ref{Fig_L_P_m} (right-hand panel) shows the $P_\mathrm{CR}/P_\mathrm{rad}$ ratio as a function of the accretion rate $\dot{m}$ for different values of the CR diffusion coefficient ($\kappa_\mathrm{CR}$). 
We observe that the $P_\mathrm{CR}/P_\mathrm{rad}$ ratio is always greater than unity for $\dot{m} < \dot{m}_c$, and is larger for smaller CR diffusivities. 
In physical terms, a smaller CR diffusion coefficient (associated with enhanced CR scattering rate) ensures that the CRs are efficiently confined, whereas CRs tend to easily escape for large diffusivities. 


\section{Cosmic rays-dust coupling in the circum-galactic medium}
\label{Sect_CR-dust_CGM}

Recent observations have revealed the existence of significant amounts of cool gas and large dust reservoirs in the circumgalactic medium (CGM) \citep[][and references therein]{Tumlinson_et_2017}. This requires some form of transport mechanism whereby the heavy elements produced by stellar evolution inside galaxies can be ejected on CGM scales. Galactic outflows provide a natural way of transporting metal-enriched gas from the galaxy to the surrounding CGM, and different detection techniques -- emission, absorption, scattering -- can be used to constrain the outflow properties \citep{Hodges-Kluck_et_2019}. Significant dust reservoirs are observed on CGM scales \citep{Peeples_et_2014, Peek_et_2015}, although the actual amount of dust present in the CGM is not a critical factor here, since CR-driving takes over radiation pressure-driving on large scales.

In the previous sections, we have seen that galactic outflows at large radii and low accretion rates are preferentially launched by CRs. Numerical simulations suggest that CR-driven outflows can have a great impact on large scales in the CGM \citep{Hopkins_et_2021}. But as the CRs propagate in the low-density environment of the CGM, the CR diffusivity tends to increase (e.g. $\kappa_\mathrm{CR} \propto \rho^{-1/4}$). This is because the CR scattering rate decreases in the low-density CGM, such that the CR diffusion coefficient increases with galactocentric distance. The enhanced CR diffusivity implies a reduced CR pressure in the CGM, and hence less efficient outflow driving. In fact, the corresponding CR diffusion timescale $t_\mathrm{diff} \sim H/\kappa_\mathrm{CR}$ becomes shorter for larger diffusivities. As a consequence, CRs tend to easily escape rather than being efficiently confined within the CGM. 

Here comes into play the CR-dust coupling. In fact, the transport of CRs can be significantly affected by interactions with dust grains present in the ambient medium. This occurs because the small-scale Alfven waves interact with both $\sim$GeV protons (dominating the CR energy density) and charged dust grains. Dust grains accelerated by radiation pressure and moving at super-Alfvenic speeds can excite small-scale Alfven waves that efficiently scatter the CRs and thus enhance the CR confinement in the CGM \citep{Squire_et_2021}. Such `dust-enhanced CR confinement' may in turn favour the development of CR-driven outflows in the CGM. Therefore radiation pressure on dust may also have a further effect in indirectly sustaining galactic outflows powered by CRs. 

Dusty outflows, driven by radiation pressure on dust, propagate to large radii and contribute to the metal enrichment of the CGM. We have previously discussed how outflowing dusty shells can reach radii of a few tens to hundred of kiloparsecs within typical AGN activity timescales \citep{Ishibashi_Fabian_2016a}. Furthermore, higher dust-to-gas ratios imply that the more dusty gas is more easily ejected by radiative feedback. Such preferential removal of dusty gas may naturally account for the presence of dust reservoirs observed in the CGM \citep{Peeples_et_2014, Peek_et_2015}. In the most extreme cases, metal-enriched gas may be transported out to 100 kiloparsec-scales by huge galactic outflows, as observed in the Makani galaxy \citep{Rupke_et_2019}.

In our picture, dusty gas can be efficiently accelerated by radiation pressure in the high-accretion regime. 
The resulting dust grains, distributed on large scales, may then help confine the CRs in the CGM and support the development of CR-driven outflows in the low-accretion regime. We remark that it is the dust that is moving through the gas with super-Alfvenic velocities. Small dust grains (with sizes $a_d \lesssim 0.01 \mu m$) seem to be required to efficiently scatter the $\sim$GeV protons and thus enable efficient CR confinement \citep{Squire_et_2021}. There could also be a dependence on the metallicity of the ambient medium, with stronger CR confinement expected at high metallicities. 

As mentioned in the Introduction, there is ample observational evidence for AGN jet-mode feedback in the form of jet-inflated bubbles in cool core clusters \citep[][and references therein]{Fabian_2012}. The radio-emitting bubbles likely contain CR protons, which may be the dominant component in e.g. Fanaroff-Riley type I lobes \citep{Croston_et_2018}. Numerical simulations show that CR-dominated light jets can naturally produce wide and fat cavities, similar to the quasi-spherical X-ray cavities (radio bubbles) present in the cores of clusters \citep{Guo_Mathews_2011, Yang_et_2019}. Such bubbles of relativistic plasma, powered by the central AGN jet, can successively expand and push out the surrounding hot gas, as observed in clusters of galaxies.  
A similar process might also operate on reduced scales in the CGM. 

The subsequent evolution of the CR-filled bubbles in the weakly magnetised CGM/ICM will depend on the fluid instabilities and magnetic interactions. Magnetic draping \citep[e.g.][]{Lyutikov_2006} can suppress the interface instabilities, such as Kelvin-Helmholtz and Rayleigh-Taylor instabilities, and initially confine the CRs inside the bubbles \citep{Ehlert_et_2018}. Eventually the CRs escape and mix with the surrounding ICM, transferring their energy to the intracluster gas \citep{Ruszkowski_et_2017}. This may provide another ICM heating mechanism that contributes to offset the cooling flows in galaxy clusters. 


\section{ Discussion } 
\label{Sect_Discussion}

\subsection{AGN cool feedback by analogy with XRB}

CRs are currently gaining interest as a potential feedback mechanism for powering cool outflows in galaxies. The viability of CRs as an efficient feedback mechanism in galaxies depends on the CR diffusion coefficient, which is a key parameter governing CR transport. It is now broadly agreed that CR feedback can drive cool ($T \lesssim 10^4$ K) outflows that also tend to be smoother and lower-speed compared to thermally-driven winds \citep{Girichidis_et_2018, Hopkins_et_2021, Farcy_et_2022}. 

Many studies in the literature consider CRs produced by supernovae in disc galaxies, whereby the CR properties are tied to the star formation properties (e.g. star formation rate). We instead consider CRs accelerated in the radio jets of AGNs. In this case, the CR characteristics are linked to the central black hole properties, with the CR injection rate scaling with the jet power. The strength of the AGN jet may in turn depend on the underlying accretion state by analogy with Galactic XRBs \citep{Churazov_et_2005}. 

Recent studies comparing different AGN classes to XRB spectral states indicate that low-luminosity radio-loud AGN could be analogs of the XRB hard state, while luminous radio-quiet AGN may represent counterparts of the XRB soft state. For instance, the correlation between the UV-to-X-ray spectral index ($\alpha_\mathrm{OX}$) and the Eddington ratio in AGNs is found to be very similar to state transitions in XRB outbursts \citep{Ruan_et_2019}. The relationship between the accretion disc and corona luminosities in radiatively efficient radio-quiet AGNs also seems to be comparable to that of the soft-intermediate state in the prototype X-ray binary GX 339-4 \citep{Arcodia_et_2020}. Most recently, X-ray observations of low-luminosity AGNs indicate a transition around a critical Eddington ratio of $\gtrsim 10^{-3}$, suggestive of a change in the underlying accretion flow modes, again similar to what observed in XRB \citep{Diaz_et_2022}. 

Furthermore, radio-loud AGNs are found to be located in distinct regions of the hardness-intensity diagram -- depending on the radio jet morphology and excitation class -- akin to XRBs \citep{Moravec_et_2022}. An analogous luminosity-excitation diagram, based on mid-infrared nebular lines, has also been proposed for AGNs. A detailed analysis of a sample of Seyfert galaxies and LINERs indicate that the characteristic `q-shape' pattern of XRB outburst cycles can be reproduced by the distribution of AGNs in such diagrams \citep{Fernandez-Ontiveros_Munoz-Darias_2021}. All these works corroborate previous results on the scale-invariance of accreting black holes (cf. Introduction), suggesting that different AGN accretion regimes may correspond to distinct XRB spectral states. 

By analogy with Galactic XRBs, we assume that the AGN jet is dominant at low accretion rates (hard state), whereas the radiative output dominates at high accretion rates (soft state). As a result, CR-driven outflows prevail in the low-accretion regime, while radiation-driven outflows take over in the high-accretion regime. CR-powered outflows are more easily launched at low accretion rates and at large radii, a regime where radiation pressure-driving is inefficient. In fact, radiation-driven outflows require relatively high accretion rates (at significant fractions of the Eddington rate) and are facilitated by IR radiation trapping in the nuclear regions. The combination of the accretion and radial dependences suggest that the two feedback mechanisms may play complementary roles in launching cool outflows on galactic scales. 

In addition, dusty outflows driven by radiation pressure are responsible for transporting and spreading dust grains into the surrounding CGM. The resulting CR-dust coupling helps confine the CRs that otherwise tend to escape (dust-enhanced CR confinement). Indeed, at high accretion rates, radiation pressure-driven outflows efficiently accelerate the dust grains on CGM scales, which then help the development of CR-driven outflows at low accretion rates. Therefore, the two feedback mechanisms not only complement each other but could also play supporting roles, with radiation pressure indirectly sustaining CR-driven outflows on large scales.  

Overall, the accretion-dependent outflow driving mechanism may suggest the following evolutionary picture.  
At early times, high accretion rates lead to powerful radiation pressure-driven outflows on galactic scales, which also transport dust into the CGM. At later times, following the decay in the accretion rate, large-scale CR-driven outflows can develop sustained by dust-enhanced CR confinement in the CGM. A two-stage feedback process could then be envisaged, with the transition from radiation pressure- to CR-driven outflows, likely corresponding to a transition in the underlying accretion flow modes. The earlier phase of radiatively efficient accretion may be associated with quasar-mode feedback at high redshifts, while the later radiatively inefficient stage may be linked to jet-mode or maintenance-mode feedback in the local Universe. 


\subsection{ Model limitations and future outlook } 

A number of physical processes are neglected and some simplifying assumptions are adopted in our analytic study. First of all, we assume CR feedback in the diffusion limit, without explicitly considering CR streaming. In more realistic situations, both diffusion and streaming of CRs should contribute to CR transport \citep{Wiener_et_2017, Farber_et_2018, Hopkins_et_2021_b, Quataert_et_2022_b}. The relative importance of CR diffusion vs. CR streaming has been investigated in a series of galaxy simulations \citep{Wiener_et_2017}. Their results suggest that transport dominated by CR diffusion can drive stronger winds compared to those driven by CR streaming. 

More recently, \cite{Quataert_et_2022_b} show that diffusive CR transport lead to more powerful winds -- with higher terminal speeds and larger mass loss rates -- than streaming transport. CR streaming is therefore inefficient at directly driving cool outflows from galaxies. Moreover, they explicitly show that the inclusion of even a small diffusion component in streaming-only models can significantly affect the resulting outflow properties; whereas including CR streaming in diffusion-dominated models has little effect. These results support our choice of focusing on CR diffusion as a promising mechanism for driving cool outflows on galactic scales, parametrised by the diffusion coefficient $\kappa_\mathrm{CR}$. 

In some previous works, $\kappa_\mathrm{CR}$ has been interpreted as an `effective' diffusion coefficient 
$\kappa_\mathrm{CR,A} \sim l_\mathrm{CR} v_A$ (where $l_\mathrm{CR}$ is the CR gradient length scale and $v_A$ is the Alfven velocity) including the combined effect of spatial diffusion and streaming \citep{Sharma_et_2009, Ehlert_et_2018}. In reality, the actual CR diffusivity is determined by the local plasma conditions as well as CR energy \citep[][]{Girichidis_et_2022}. For instance, CRs tend to propagate faster in a cold neutral medium than in hot ionised gas, implying an enhanced $\kappa_\mathrm{CR}$ in the first case. An improved treatment may be to adopt a temperature-dependent CR diffusion coefficient $\kappa_\mathrm{CR}(T)$ to mimic the different coupling of CRs in different gas-phase environments \citep{Farber_et_2018}. 

Another process that needs to be considered is the collisional loss mechanism for the CR population: CR particles lose energy to the surrounding thermal gas via Coulomb and hadronic interactions \citep{Pfrommer_et_2017}.
Low-energy CR protons mostly lose energy via Coulomb collisions with thermal gas, while high-energy CR particles interact inelastically with the nuclei of the thermal plasma. As a result, CRs in the low-momentum regime (with momenta $\lesssim \mathrm{GeV/c}$) efficiently cool via Coulomb losses and eventually get thermalised. Conversely, hadronic losses dominate in the high-momentum regime ($\gtrsim \mathrm{GeV/c}$), leading to the production of neutral pions, which subsequently decay into $\gamma$-ray photons \citep{Pfrommer_et_2017, Wiener_et_2017, Chan_et_2019}. As a consequence, CRs can suffer strong hadronic losses in high-density environments, as the loss is proportional to the gas density and CR energy density \citep{Crocker_et_2021_b}. 

Furthermore, CRs can undergo streaming losses due to the excitation of Alfven waves and subsequent wave damping \citep{Ehlert_et_2018, Chan_et_2019, Hopkins_et_2020}. This process removes energy from the CR population and heat the surrounding gas, resulting in Alfven wave cooling for the CRs and Alfven wave heating for the gas, respectively. Such Alfven waves may be responsible for the difference observed in the wind properties between diffusion vs. streaming models \citep{Wiener_et_2017}. Ideally, we should consider both diffusion and streaming for CR transport, and the different loss mechanisms (Coulomb, hadronic, and streaming) should be taken into account in future models of CR feedback. 

Concerning radiation feedback, an important requirement in our model is the presence of dust, since radiation pressure mainly acts on the dust grains. A constant dust-to-gas ratio is implicitly assumed throughout the outflow propagation, which is likely an optimistic assumption. Dust grains can be destroyed by thermal/non-thermal sputtering and in shocks, leading to dust depletion on galactic scales \citep[][and references therein]{Veilleux_et_2020}. At high temperatures, the thermal sputtering of dust grains can rapidly reduce the dust mass at large galactocentric radii \citep{Barnes_et_2020}. In our picture, the dusty gas being driven out by radiation pressure is strongly radiating and remains cool (we recall that the radiative cooling of dense gas is very efficient). The evolution of dusty gas could be affected by  a number of other processes, such as photo-evaporation due to far-UV radiation, thermal conduction, Compton heating and X-ray emission \citep{Ferrara_Scannapieco_2016, Bruggen_Scannapieco_2016, Xie_et_2017}\footnote{Actually, X-ray emission from the central AGN allows the gas and dust particles to stay weakly ionised, so that they are efficiently coupled by Coulomb forces \citep{Fabian_et_2008}.}. 

In addition to the different dust destruction mechanisms operating in galaxies, dust can also be created in supernova explosions. Observations indicate that large amounts of dust are efficiently produced in core-collapse supernovae, yielding high dust-to-gas ratios \citep{Owen_Barlow_2015, Wesson_et_2015, Slavin_et_2020}. Fresh dust could also be released at large radii by supernova explosions of massive stars formed inside the radiation pressure-driven outflows [e.g. see the AGN feedback-driven star formation scenario in \citet{Ishibashi_Fabian_2012} and observations of star formation within galactic outflows in \citet{Maiolino_et_2017}]. 

The strength of the radiation-matter coupling is another key aspect of the AGN radiation feedback scenario.
Here we assume a spherically symmetric thin shell with a homogenous gas distribution \citep{Thompson_et_2015, Ishibashi_Fabian_2015}. Recent RHD simulations of radiation-driven shells indicate that the boost factor is roughly equal to the IR optical depth -- at least for moderate optical depths -- in agreement with simple analytic predictions \citep{Costa_et_2018}. But as the shell expands outwards, it can be disrupted and fragmented into multiple clouds, giving rise to a clumpy  distribution. The reprocessed IR photons may leak out through lower density channels, reducing the effective radiation-matter coupling efficiency. The number of IR multi-scatterings could then be lowered to roughly one-quarter of the IR optical depth, as reported in RHD simulations with inhomogeneous interstellar medium \citep{Bieri_et_2017}. It is worth mentioning that the issue of radiation trapping and fluid instabilities has been investigated in different numerical simulations, with somewhat contrasting results reported in the literature \citep{Krumholz_Thompson_2013, Davis_et_2014, Tsang_Milosavljevic_2015, Zhang_Davis_2017}.  

In future, our AGN feedback model could be generalised to non-spherical geometries, including a more realistic clumpy gas distribution. This should also allow us to perform more detailed one-by-one comparisons with observations of galactic outflows \citep{Ishibashi_et_2021}. While we now focus on the outflow launching conditions, in subsequent studies we wish to self-consistently follow the evolution of CR and radiation pressure-driven outflows in the different accretion regimes over cosmic time. 


\section{ Conclusion } 
\label{Sect_Conclusion}

We compare the relative importance of CRs and radiation pressure on dust in powering cool outflows in galaxies. 
The CR Eddington limit can be more easily exceeded at larger radii compared to the radiative Eddington limit. The minimum accretion rate required for outflow driving is lower for CRs than for radiation pressure on dust. Assuming that CRs originate in AGN jets and by analogy with the XRB spectral states, we analyse the accretion-dependent outflow driving mechanisms. We obtain that CR-driven outflows can be easily launched at low accretion rates and large radii, precisely where radiation pressure-driving becomes inefficient. Conversely, radiation pressure-driven outflows dominate at high accretion rates and small radii (where radiation trapping can be important). Therefore the two AGN feedback mechanisms seem to complement each other in driving cool outflows on galactic scales.  

In addition, dusty outflows driven by radiation pressure in the high-accretion regime can transport dust grains into the surrounding CGM. The resulting `dust-enhanced CR confinement' can further support the development of large-scale CR-driven outflows in the low-accretion regime. The transition from radiation pressure-driven outflows at early times to CR-driven outflows at later times may reflect the transition in the underlying accretion flow modes (from radiatively efficient accretion disc to radiatively inefficient jet-dominated flow). In conclusion, the two AGN feedback mechanisms --- radiation and CR pressure --- may play both complementary and supporting roles in driving cool galactic outflows, one within the host galaxy, the other within the CGM.


\section*{Data availability}

No new data were generated or analysed in support of this research.


\bibliographystyle{mn2e}
\bibliography{biblio.bib}

\begin{thebibliography}{}

\bibitem[\protect\citeauthoryear{{Arcodia}, {Ponti}, {Merloni} \&
  {Nandra}}{{Arcodia} et~al.}{2020}]{Arcodia_et_2020}
{Arcodia} R.,  {Ponti} G.,  {Merloni} A.,    {Nandra} K.,  2020, \aap, 638,
  A100

\bibitem[\protect\citeauthoryear{{Barnes}, {Kannan}, {Vogelsberger} \&
  {Marinacci}}{{Barnes} et~al.}{2020}]{Barnes_et_2020}
{Barnes} D.~J.,  {Kannan} R.,  {Vogelsberger} M.,    {Marinacci} F.,  2020,
  \mnras, 494, 1143

\bibitem[\protect\citeauthoryear{{Bieri}, {Dubois}, {Rosdahl}, {Wagner}, {Silk}
  \& {Mamon}}{{Bieri} et~al.}{2017}]{Bieri_et_2017}
{Bieri} R.,  {Dubois} Y.,  {Rosdahl} J.,  {Wagner} A.,  {Silk} J.,    {Mamon}
  G.~A.,  2017, \mnras, 464, 1854

\bibitem[\protect\citeauthoryear{{B{\^\i}rzan}, {Rafferty}, {Br{\"u}ggen},
  {Botteon}, {Brunetti}, {Cuciti}, {Edge}, {Morganti}, {R{\"o}ttgering} \&
  {Shimwell}}{{B{\^\i}rzan} et~al.}{2020}]{Birzan_et_2020}
{B{\^\i}rzan} L.,  {Rafferty} D.~A.,  {Br{\"u}ggen} M.,  {Botteon} A.,
  {Brunetti} G.,  {Cuciti} V.,  {Edge} A.~C.,  {Morganti} R.,  {R{\"o}ttgering}
  H.~J.~A.,    {Shimwell} T.~W.,  2020, \mnras, 496, 2613

\bibitem[\protect\citeauthoryear{{Br{\"u}ggen} \& {Scannapieco}}{{Br{\"u}ggen}
  \& {Scannapieco}}{2016}]{Bruggen_Scannapieco_2016}
{Br{\"u}ggen} M.,  {Scannapieco} E.,  2016, \apj, 822, 31

\bibitem[\protect\citeauthoryear{{Chan}, {Kere{\v{s}}}, {Hopkins}, {Quataert},
  {Su}, {Hayward} \& {Faucher-Gigu{\`e}re}}{{Chan} et~al.}{2019}]{Chan_et_2019}
{Chan} T.~K.,  {Kere{\v{s}}} D.,  {Hopkins} P.~F.,  {Quataert} E.,  {Su} K.~Y.,
   {Hayward} C.~C.,    {Faucher-Gigu{\`e}re} C.~A.,  2019, \mnras, 488, 3716

\bibitem[\protect\citeauthoryear{{Churazov}, {Sazonov}, {Sunyaev}, {Forman},
  {Jones} \& {B{\"o}hringer}}{{Churazov} et~al.}{2005}]{Churazov_et_2005}
{Churazov} E.,  {Sazonov} S.,  {Sunyaev} R.,  {Forman} W.,  {Jones} C.,
  {B{\"o}hringer} H.,  2005, \mnras, 363, L91

\bibitem[\protect\citeauthoryear{{Costa}, {Rosdahl}, {Sijacki} \&
  {Haehnelt}}{{Costa} et~al.}{2018}]{Costa_et_2018}
{Costa} T.,  {Rosdahl} J.,  {Sijacki} D.,    {Haehnelt} M.~G.,  2018, \mnras,
  473, 4197

\bibitem[\protect\citeauthoryear{{Crocker}, {Krumholz} \& {Thompson}}{{Crocker}
  et~al.}{2021}]{Crocker_et_2021_b}
{Crocker} R.~M.,  {Krumholz} M.~R.,    {Thompson} T.~A.,  2021, \mnras, 503,
  2651

\bibitem[\protect\citeauthoryear{{Croston}, {Ineson} \& {Hardcastle}}{{Croston}
  et~al.}{2018}]{Croston_et_2018}
{Croston} J.~H.,  {Ineson} J.,    {Hardcastle} M.~J.,  2018, \mnras, 476, 1614

\bibitem[\protect\citeauthoryear{{Davis}, {Jiang}, {Stone} \& {Murray}}{{Davis}
  et~al.}{2014}]{Davis_et_2014}
{Davis} S.~W.,  {Jiang} Y.-F.,  {Stone} J.~M.,    {Murray} N.,  2014, \apj,
  796, 107

\bibitem[\protect\citeauthoryear{{D{\'\i}az}, {Hern{\'a}ndez-Garc{\'\i}a},
  {Ar{\'e}valo}, {L{\'o}pez-Navas}, {Ricci}, {Koss},
  {Gonz{\'a}lez-Mart{\'\i}n}, {Balokovi{\'c}}, {Osorio-Clavijo}, {Garc{\'\i}a}
  \& {Malizia}}{{D{\'\i}az} et~al.}{2022}]{Diaz_et_2022}
{D{\'\i}az} Y.,  {Hern{\'a}ndez-Garc{\'\i}a} L.,  {Ar{\'e}valo} P.,
  {L{\'o}pez-Navas} E.,  {Ricci} C.,  {Koss} M.,  {Gonz{\'a}lez-Mart{\'\i}n}
  O.,  {Balokovi{\'c}} M.,  {Osorio-Clavijo} N.,  {Garc{\'\i}a} J.,
  {Malizia} A.,  2022, arXiv e-prints, p. arXiv:2210.15376

\bibitem[\protect\citeauthoryear{{Dunn} \& {Fabian}}{{Dunn} \&
  {Fabian}}{2004}]{Dunn_Fabian_2004}
{Dunn} R.~J.~H.,  {Fabian} A.~C.,  2004, \mnras, 355, 862

\bibitem[\protect\citeauthoryear{{Ehlert}, {Weinberger}, {Pfrommer}, {Pakmor}
  \& {Springel}}{{Ehlert} et~al.}{2018}]{Ehlert_et_2018}
{Ehlert} K.,  {Weinberger} R.,  {Pfrommer} C.,  {Pakmor} R.,    {Springel} V.,
  2018, \mnras, 481, 2878

\bibitem[\protect\citeauthoryear{{Fabian}}{{Fabian}}{1999}]{Fabian_1999}
{Fabian} A.~C.,  1999, \mnras, 308, L39

\bibitem[\protect\citeauthoryear{{Fabian}}{{Fabian}}{2012}]{Fabian_2012}
{Fabian} A.~C.,  2012, \araa, 50, 455

\bibitem[\protect\citeauthoryear{{Fabian}, {Vasudevan} \& {Gandhi}}{{Fabian}
  et~al.}{2008}]{Fabian_et_2008}
{Fabian} A.~C.,  {Vasudevan} R.~V.,    {Gandhi} P.,  2008, \mnras, 385, L43

\bibitem[\protect\citeauthoryear{{Falcke}, {K{\"o}rding} \& {Markoff}}{{Falcke}
  et~al.}{2004}]{Falcke_et_2004}
{Falcke} H.,  {K{\"o}rding} E.,    {Markoff} S.,  2004, \aap, 414, 895

\bibitem[\protect\citeauthoryear{{Farber}, {Ruszkowski}, {Yang} \&
  {Zweibel}}{{Farber} et~al.}{2018}]{Farber_et_2018}
{Farber} R.,  {Ruszkowski} M.,  {Yang} H. Y.~K.,    {Zweibel} E.~G.,  2018,
  \apj, 856, 112

\bibitem[\protect\citeauthoryear{{Farcy}, {Rosdahl}, {Dubois}, {Blaizot} \&
  {Martin-Alvarez}}{{Farcy} et~al.}{2022}]{Farcy_et_2022}
{Farcy} M.,  {Rosdahl} J.,  {Dubois} Y.,  {Blaizot} J.,    {Martin-Alvarez} S.,
   2022, \mnras, 513, 5000

\bibitem[\protect\citeauthoryear{{Fender}}{{Fender}}{2010}]{Fender_2010}
{Fender} R.,  2010, in {Belloni} T.,  ed., , Vol.~794, Lecture Notes in
  Physics, Berlin Springer Verlag.
p.~115

\bibitem[\protect\citeauthoryear{{Fender} \& {Mu{\~n}oz-Darias}}{{Fender} \&
  {Mu{\~n}oz-Darias}}{2016}]{Fender_Munoz_2016}
{Fender} R.,  {Mu{\~n}oz-Darias} T.,  2016, in {Haardt} F.,  {Gorini} V.,
  {Moschella} U.,  {Treves} A.,   {Colpi} M.,  eds, , Vol.~905, Lecture Notes
  in Physics, Berlin Springer Verlag.
p.~65

\bibitem[\protect\citeauthoryear{{Fender}, {Belloni} \& {Gallo}}{{Fender}
  et~al.}{2004}]{Fender_et_2004}
{Fender} R.~P.,  {Belloni} T.~M.,    {Gallo} E.,  2004, \mnras, 355, 1105

\bibitem[\protect\citeauthoryear{{Fern{\'a}ndez-Ontiveros} \&
  {Mu{\~n}oz-Darias}}{{Fern{\'a}ndez-Ontiveros} \&
  {Mu{\~n}oz-Darias}}{2021}]{Fernandez-Ontiveros_Munoz-Darias_2021}
{Fern{\'a}ndez-Ontiveros} J.~A.,  {Mu{\~n}oz-Darias} T.,  2021, \mnras, 504,
  5726

\bibitem[\protect\citeauthoryear{{Ferrara} \& {Scannapieco}}{{Ferrara} \&
  {Scannapieco}}{2016}]{Ferrara_Scannapieco_2016}
{Ferrara} A.,  {Scannapieco} E.,  2016, \apj, 833, 46

\bibitem[\protect\citeauthoryear{{Fiore}, {Feruglio}, {Shankar}, {Bischetti},
  {Bongiorno}, {Brusa}, {Carniani}, {Cicone}, {Duras}, {Lamastra}, {Mainieri},
  {Marconi}, {Menci}, {Maiolino}, {Piconcelli}, {Vietri} \&
  {Zappacosta}}{{Fiore} et~al.}{2017}]{Fiore_et_2017}
{Fiore} F.,  {Feruglio} C.,  {Shankar} F.,  {Bischetti} M.,  {Bongiorno} A.,
  {Brusa} M.,  {Carniani} S.,  {Cicone} C.,  {Duras} F.,  {Lamastra} A.,
  {Mainieri} V.,  {Marconi} A.,  {Menci} N.,  {Maiolino} R.,  {Piconcelli} E.,
  {Vietri} G.,    {Zappacosta} L.,  2017, \aap, 601, A143

\bibitem[\protect\citeauthoryear{{Fluetsch}, {Maiolino}, {Carniani}, {Arribas},
  {Belfiore}, {Bellocchi}, {Cazzoli}, {Cicone}, {Cresci}, {Fabian},
  {Gallagher}, {Ishibashi}, {Mannucci}, {Marconi}, {Perna}, {Sturm} \&
  {Venturi}}{{Fluetsch} et~al.}{2020}]{Fluetsch_et_2020}
{Fluetsch} A.,  {Maiolino} R.,  {Carniani} S.,  {Arribas} S.,  {Belfiore} F.,
  {Bellocchi} E.,  {Cazzoli} S.,  {Cicone} C.,  {Cresci} G.,  {Fabian} A.~C.,
  {Gallagher} R.,  {Ishibashi} W.,  {Mannucci} F.,  {Marconi} A.,  {Perna} M.,
  {Sturm} E.,    {Venturi} G.,  2020, arXiv e-prints, p. arXiv:2006.13232

\bibitem[\protect\citeauthoryear{{Fluetsch}, {Maiolino}, {Carniani}, {Marconi},
  {Cicone}, {Bourne}, {Costa}, {Fabian}, {Ishibashi} \& {Venturi}}{{Fluetsch}
  et~al.}{2019}]{Fluetsch_et_2019}
{Fluetsch} A.,  {Maiolino} R.,  {Carniani} S.,  {Marconi} A.,  {Cicone} C.,
  {Bourne} M.~A.,  {Costa} T.,  {Fabian} A.~C.,  {Ishibashi} W.,    {Venturi}
  G.,  2019, \mnras, 483, 4586

\bibitem[\protect\citeauthoryear{{Gallo}, {Fender}, {Kaiser}, {Russell},
  {Morganti}, {Oosterloo} \& {Heinz}}{{Gallo} et~al.}{2005}]{Gallo_et_2005}
{Gallo} E.,  {Fender} R.,  {Kaiser} C.,  {Russell} D.,  {Morganti} R.,
  {Oosterloo} T.,    {Heinz} S.,  2005, \nat, 436, 819

\bibitem[\protect\citeauthoryear{{Girichidis}, {Naab}, {Hanasz} \&
  {Walch}}{{Girichidis} et~al.}{2018}]{Girichidis_et_2018}
{Girichidis} P.,  {Naab} T.,  {Hanasz} M.,    {Walch} S.,  2018, \mnras, 479,
  3042

\bibitem[\protect\citeauthoryear{{Girichidis}, {Pfrommer}, {Pakmor} \&
  {Springel}}{{Girichidis} et~al.}{2022}]{Girichidis_et_2022}
{Girichidis} P.,  {Pfrommer} C.,  {Pakmor} R.,    {Springel} V.,  2022, \mnras,
  510, 3917

\bibitem[\protect\citeauthoryear{{Guo} \& {Mathews}}{{Guo} \&
  {Mathews}}{2011}]{Guo_Mathews_2011}
{Guo} F.,  {Mathews} W.~G.,  2011, \apj, 728, 121

\bibitem[\protect\citeauthoryear{{Heckman} \& {Thompson}}{{Heckman} \&
  {Thompson}}{2017}]{Heckman_Thompson_2017}
{Heckman} T.~M.,  {Thompson} T.~A.,  2017, arXiv e-prints, p. arXiv:1701.09062

\bibitem[\protect\citeauthoryear{{Heintz} \& {Zweibel}}{{Heintz} \&
  {Zweibel}}{2022}]{Heintz_Zweibel_2022}
{Heintz} E.,  {Zweibel} E.,  2022, arXiv e-prints, p. arXiv:2206.04082

\bibitem[\protect\citeauthoryear{{Hodges-Kluck}, {Corrales}, {Veilleux},
  {Bregman}, {Li} \& {Melendez}}{{Hodges-Kluck}
  et~al.}{2019}]{Hodges-Kluck_et_2019}
{Hodges-Kluck} E.,  {Corrales} L.,  {Veilleux} S.,  {Bregman} J.,  {Li} J.,
  {Melendez} M.,  2019, \baas, 51, 249

\bibitem[\protect\citeauthoryear{{Hopkins}, {Chan}, {Garrison-Kimmel}, {Ji},
  {Su}, {Hummels}, {Kere{\v{s}}}, {Quataert} \&
  {Faucher-Gigu{\`e}re}}{{Hopkins} et~al.}{2020}]{Hopkins_et_2020}
{Hopkins} P.~F.,  {Chan} T.~K.,  {Garrison-Kimmel} S.,  {Ji} S.,  {Su} K.-Y.,
  {Hummels} C.~B.,  {Kere{\v{s}}} D.,  {Quataert} E.,    {Faucher-Gigu{\`e}re}
  C.-A.,  2020, \mnras, 492, 3465

\bibitem[\protect\citeauthoryear{{Hopkins}, {Chan}, {Ji}, {Hummels},
  {Kere{\v{s}}}, {Quataert} \& {Faucher-Gigu{\`e}re}}{{Hopkins}
  et~al.}{2021}]{Hopkins_et_2021}
{Hopkins} P.~F.,  {Chan} T.~K.,  {Ji} S.,  {Hummels} C.~B.,  {Kere{\v{s}}} D.,
  {Quataert} E.,    {Faucher-Gigu{\`e}re} C.-A.,  2021, \mnras, 501, 3640

\bibitem[\protect\citeauthoryear{{Hopkins}, {Chan}, {Squire}, {Quataert}, {Ji},
  {Kere{\v{s}}} \& {Faucher-Gigu{\`e}re}}{{Hopkins}
  et~al.}{2021}]{Hopkins_et_2021_b}
{Hopkins} P.~F.,  {Chan} T.~K.,  {Squire} J.,  {Quataert} E.,  {Ji} S.,
  {Kere{\v{s}}} D.,    {Faucher-Gigu{\`e}re} C.-A.,  2021, \mnras, 501, 3663

\bibitem[\protect\citeauthoryear{{Huang} \& {Davis}}{{Huang} \&
  {Davis}}{2022}]{Huang_Davis_2022}
{Huang} X.,  {Davis} S.~W.,  2022, \mnras, 511, 5125

\bibitem[\protect\citeauthoryear{{Ipavich}}{{Ipavich}}{1975}]{Ipavich_1975}
{Ipavich} F.~M.,  1975, \apj, 196, 107

\bibitem[\protect\citeauthoryear{{Ishibashi} \& {Fabian}}{{Ishibashi} \&
  {Fabian}}{2012}]{Ishibashi_Fabian_2012}
{Ishibashi} W.,  {Fabian} A.~C.,  2012, \mnras, 427, 2998

\bibitem[\protect\citeauthoryear{{Ishibashi} \& {Fabian}}{{Ishibashi} \&
  {Fabian}}{2015}]{Ishibashi_Fabian_2015}
{Ishibashi} W.,  {Fabian} A.~C.,  2015, \mnras, 451, 93

\bibitem[\protect\citeauthoryear{{Ishibashi} \& {Fabian}}{{Ishibashi} \&
  {Fabian}}{2016a}]{Ishibashi_Fabian_2016b}
{Ishibashi} W.,  {Fabian} A.~C.,  2016a, \mnras, 463, 1291

\bibitem[\protect\citeauthoryear{{Ishibashi} \& {Fabian}}{{Ishibashi} \&
  {Fabian}}{2016b}]{Ishibashi_Fabian_2016a}
{Ishibashi} W.,  {Fabian} A.~C.,  2016b, \mnras, 457, 2864

\bibitem[\protect\citeauthoryear{{Ishibashi}, {Fabian} \&
  {Arakawa}}{{Ishibashi} et~al.}{2021}]{Ishibashi_et_2021}
{Ishibashi} W.,  {Fabian} A.~C.,    {Arakawa} N.,  2021, \mnras, 502, 3638

\bibitem[\protect\citeauthoryear{{Ishibashi}, {Fabian} \&
  {Maiolino}}{{Ishibashi} et~al.}{2018}]{Ishibashi_et_2018a}
{Ishibashi} W.,  {Fabian} A.~C.,    {Maiolino} R.,  2018, \mnras, 476, 512

\bibitem[\protect\citeauthoryear{{K{\"o}rding}, {Jester} \&
  {Fender}}{{K{\"o}rding} et~al.}{2006}]{Koerding_et_2006}
{K{\"o}rding} E.~G.,  {Jester} S.,    {Fender} R.,  2006, \mnras, 372, 1366

\bibitem[\protect\citeauthoryear{{K{\"o}rding}, {Jester} \&
  {Fender}}{{K{\"o}rding} et~al.}{2008}]{Koerding_et_2008}
{K{\"o}rding} E.~G.,  {Jester} S.,    {Fender} R.,  2008, \mnras, 383, 277

\bibitem[\protect\citeauthoryear{{Kormendy} \& {Ho}}{{Kormendy} \&
  {Ho}}{2013}]{Kormendy_Ho_2013}
{Kormendy} J.,  {Ho} L.~C.,  2013, \araa, 51, 511

\bibitem[\protect\citeauthoryear{{Krumholz} \& {Thompson}}{{Krumholz} \&
  {Thompson}}{2013}]{Krumholz_Thompson_2013}
{Krumholz} M.~R.,  {Thompson} T.~A.,  2013, \mnras, 434, 2329

\bibitem[\protect\citeauthoryear{{Lyutikov}}{{Lyutikov}}{2006}]{Lyutikov_2006}
{Lyutikov} M.,  2006, \mnras, 373, 73

\bibitem[\protect\citeauthoryear{{Maiolino}, {Russell}, {Fabian}, {Carniani},
  {Gallagher}, {Cazzoli}, {Arribas}, {Belfiore}, {Bellocchi}, {Colina},
  {Cresci}, {Ishibashi}, {Marconi}, {Mannucci}, {Oliva} \& {Sturm}}{{Maiolino}
  et~al.}{2017}]{Maiolino_et_2017}
{Maiolino} R.,  {Russell} H.~R.,  {Fabian} A.~C.,  {Carniani} S.,  {Gallagher}
  R.,  {Cazzoli} S.,  {Arribas} S.,  {Belfiore} F.,  {Bellocchi} E.,  {Colina}
  L.,  {Cresci} G.,  {Ishibashi} W.,  {Marconi} A.,  {Mannucci} F.,  {Oliva}
  E.,    {Sturm} E.,  2017, \nat, 544, 202

\bibitem[\protect\citeauthoryear{{McHardy}, {Koerding}, {Knigge}, {Uttley} \&
  {Fender}}{{McHardy} et~al.}{2006}]{McHardy_et_2006}
{McHardy} I.~M.,  {Koerding} E.,  {Knigge} C.,  {Uttley} P.,    {Fender} R.~P.,
   2006, \nat, 444, 730

\bibitem[\protect\citeauthoryear{{McNamara} \& {Nulsen}}{{McNamara} \&
  {Nulsen}}{2012}]{McNamara_Nulsen_2012}
{McNamara} B.~R.,  {Nulsen} P.~E.~J.,  2012, New Journal of Physics, 14, 055023

\bibitem[\protect\citeauthoryear{{Merloni} \& {Heinz}}{{Merloni} \&
  {Heinz}}{2008}]{Merloni_Heinz_2008}
{Merloni} A.,  {Heinz} S.,  2008, \mnras, 388, 1011

\bibitem[\protect\citeauthoryear{{Merloni}, {Heinz} \& {di Matteo}}{{Merloni}
  et~al.}{2003}]{Merloni_et_2003}
{Merloni} A.,  {Heinz} S.,    {di Matteo} T.,  2003, \mnras, 345, 1057

\bibitem[\protect\citeauthoryear{{Mocz}, {Fabian} \& {Blundell}}{{Mocz}
  et~al.}{2013}]{Mocz_et_2013}
{Mocz} P.,  {Fabian} A.~C.,    {Blundell} K.~M.,  2013, \mnras, 432, 3381

\bibitem[\protect\citeauthoryear{{Moravec}, {Svoboda}, {Borkar}, {Boorman},
  {Kynoch}, {Panessa}, {Mingo} \& {Guainazzi}}{{Moravec}
  et~al.}{2022}]{Moravec_et_2022}
{Moravec} E.,  {Svoboda} J.,  {Borkar} A.,  {Boorman} P.,  {Kynoch} D.,
  {Panessa} F.,  {Mingo} B.,    {Guainazzi} M.,  2022, \aap, 662, A28

\bibitem[\protect\citeauthoryear{{Murray}, {M{\'e}nard} \& {Thompson}}{{Murray}
  et~al.}{2011}]{Murray_et_2011}
{Murray} N.,  {M{\'e}nard} B.,    {Thompson} T.~A.,  2011, \apj, 735, 66

\bibitem[\protect\citeauthoryear{{Murray}, {Quataert} \& {Thompson}}{{Murray}
  et~al.}{2005}]{Murray_et_2005}
{Murray} N.,  {Quataert} E.,    {Thompson} T.~A.,  2005, \apj, 618, 569

\bibitem[\protect\citeauthoryear{{Owen} \& {Barlow}}{{Owen} \&
  {Barlow}}{2015}]{Owen_Barlow_2015}
{Owen} P.~J.,  {Barlow} M.~J.,  2015, ArXiv e-prints

\bibitem[\protect\citeauthoryear{{Peek}, {M{\'e}nard} \& {Corrales}}{{Peek}
  et~al.}{2015}]{Peek_et_2015}
{Peek} J.~E.~G.,  {M{\'e}nard} B.,    {Corrales} L.,  2015, \apj, 813, 7

\bibitem[\protect\citeauthoryear{{Peeples}, {Werk}, {Tumlinson}, {Oppenheimer},
  {Prochaska}, {Katz} \& {Weinberg}}{{Peeples} et~al.}{2014}]{Peeples_et_2014}
{Peeples} M.~S.,  {Werk} J.~K.,  {Tumlinson} J.,  {Oppenheimer} B.~D.,
  {Prochaska} J.~X.,  {Katz} N.,    {Weinberg} D.~H.,  2014, \apj, 786, 54

\bibitem[\protect\citeauthoryear{{Pfrommer}, {Pakmor}, {Schaal}, {Simpson} \&
  {Springel}}{{Pfrommer} et~al.}{2017}]{Pfrommer_et_2017}
{Pfrommer} C.,  {Pakmor} R.,  {Schaal} K.,  {Simpson} C.~M.,    {Springel} V.,
  2017, \mnras, 465, 4500

\bibitem[\protect\citeauthoryear{{Qiu}, {Bogdanovi{\'c}}, {Li}, {Park} \&
  {Wise}}{{Qiu} et~al.}{2019}]{Qiu_et_2019}
{Qiu} Y.,  {Bogdanovi{\'c}} T.,  {Li} Y.,  {Park} K.,    {Wise} J.~H.,  2019,
  \apj, 877, 47

\bibitem[\protect\citeauthoryear{{Quataert}, {Jiang} \& {Thompson}}{{Quataert}
  et~al.}{2022}]{Quataert_et_2022_b}
{Quataert} E.,  {Jiang} Y.-F.,    {Thompson} T.~A.,  2022, \mnras, 510, 920

\bibitem[\protect\citeauthoryear{{Quataert}, {Thompson} \& {Jiang}}{{Quataert}
  et~al.}{2022}]{Quataert_et_2022_a}
{Quataert} E.,  {Thompson} T.~A.,    {Jiang} Y.-F.,  2022, \mnras, 510, 1184

\bibitem[\protect\citeauthoryear{{Ruan}, {Anderson}, {Eracleous}, {Green},
  {Haggard}, {MacLeod}, {Runnoe} \& {Sobolewska}}{{Ruan}
  et~al.}{2019}]{Ruan_et_2019}
{Ruan} J.~J.,  {Anderson} S.~F.,  {Eracleous} M.,  {Green} P.~J.,  {Haggard}
  D.,  {MacLeod} C.~L.,  {Runnoe} J.~C.,    {Sobolewska} M.~A.,  2019, \apj,
  883, 76

\bibitem[\protect\citeauthoryear{{Rupke}, {Coil}, {Geach}, {Tremonti},
  {Diamond-Stanic}, {George}, {Hickox}, {Kepley}, {Leung}, {Moustakas},
  {Rudnick} \& {Sell}}{{Rupke} et~al.}{2019}]{Rupke_et_2019}
{Rupke} D. S.~N.,  {Coil} A.,  {Geach} J.~E.,  {Tremonti} C.,  {Diamond-Stanic}
  A.~M.,  {George} E.~R.,  {Hickox} R.~C.,  {Kepley} A.~A.,  {Leung} G.,
  {Moustakas} J.,  {Rudnick} G.,    {Sell} P.~H.,  2019, \nat, 574, 643

\bibitem[\protect\citeauthoryear{{Russell}, {Fender}, {Gallo} \&
  {Kaiser}}{{Russell} et~al.}{2007}]{Russell_et_2007}
{Russell} D.~M.,  {Fender} R.~P.,  {Gallo} E.,    {Kaiser} C.~R.,  2007,
  \mnras, 376, 1341

\bibitem[\protect\citeauthoryear{{Ruszkowski}, {Yang} \&
  {Reynolds}}{{Ruszkowski} et~al.}{2017}]{Ruszkowski_et_2017}
{Ruszkowski} M.,  {Yang} H. Y.~K.,    {Reynolds} C.~S.,  2017, \apj, 844, 13

\bibitem[\protect\citeauthoryear{{Sadowski} \& {Gaspari}}{{Sadowski} \&
  {Gaspari}}{2017}]{Sadowski_Gaspari_2017}
{Sadowski} A.,  {Gaspari} M.,  2017, \mnras, 468, 1398

\bibitem[\protect\citeauthoryear{{Shakura} \& {Sunyaev}}{{Shakura} \&
  {Sunyaev}}{1973}]{Shakura_Sunyaev_1973}
{Shakura} N.~I.,  {Sunyaev} R.~A.,  1973, \aap, 24, 337

\bibitem[\protect\citeauthoryear{{Sharma}, {Chandran}, {Quataert} \&
  {Parrish}}{{Sharma} et~al.}{2009}]{Sharma_et_2009}
{Sharma} P.,  {Chandran} B. D.~G.,  {Quataert} E.,    {Parrish} I.~J.,  2009,
  \apj, 699, 348

\bibitem[\protect\citeauthoryear{{Sironi} \& {Socrates}}{{Sironi} \&
  {Socrates}}{2010}]{Sironi_Socrates_2010}
{Sironi} L.,  {Socrates} A.,  2010, \apj, 710, 891

\bibitem[\protect\citeauthoryear{{Slavin}, {Dwek}, {Mac Low} \&
  {Hill}}{{Slavin} et~al.}{2020}]{Slavin_et_2020}
{Slavin} J.~D.,  {Dwek} E.,  {Mac Low} M.-M.,    {Hill} A.~S.,  2020, \apj,
  902, 135

\bibitem[\protect\citeauthoryear{{Socrates}, {Davis} \&
  {Ramirez-Ruiz}}{{Socrates} et~al.}{2008}]{Socrates_et_2008}
{Socrates} A.,  {Davis} S.~W.,    {Ramirez-Ruiz} E.,  2008, \apj, 687, 202

\bibitem[\protect\citeauthoryear{{Squire}, {Hopkins}, {Quataert} \&
  {Kempski}}{{Squire} et~al.}{2021}]{Squire_et_2021}
{Squire} J.,  {Hopkins} P.~F.,  {Quataert} E.,    {Kempski} P.,  2021, \mnras,
  502, 2630

\bibitem[\protect\citeauthoryear{{Thomas}, {Pfrommer} \& {Pakmor}}{{Thomas}
  et~al.}{2022}]{Thomas_et_2022}
{Thomas} T.,  {Pfrommer} C.,    {Pakmor} R.,  2022, arXiv e-prints, p.
  arXiv:2203.12029

\bibitem[\protect\citeauthoryear{{Thompson}, {Fabian}, {Quataert} \&
  {Murray}}{{Thompson} et~al.}{2015}]{Thompson_et_2015}
{Thompson} T.~A.,  {Fabian} A.~C.,  {Quataert} E.,    {Murray} N.,  2015,
  \mnras, 449, 147

\bibitem[\protect\citeauthoryear{{Tsang} \& {Milosavljevi{\'c}}}{{Tsang} \&
  {Milosavljevi{\'c}}}{2015}]{Tsang_Milosavljevic_2015}
{Tsang} B.~T.-H.,  {Milosavljevi{\'c}} M.,  2015, \mnras, 453, 1108

\bibitem[\protect\citeauthoryear{{Tumlinson}, {Peeples} \& {Werk}}{{Tumlinson}
  et~al.}{2017}]{Tumlinson_et_2017}
{Tumlinson} J.,  {Peeples} M.~S.,    {Werk} J.~K.,  2017, \araa, 55, 389

\bibitem[\protect\citeauthoryear{{Veilleux}, {Maiolino}, {Bolatto} \&
  {Aalto}}{{Veilleux} et~al.}{2020}]{Veilleux_et_2020}
{Veilleux} S.,  {Maiolino} R.,  {Bolatto} A.~D.,    {Aalto} S.,  2020, \aapr,
  28, 2

\bibitem[\protect\citeauthoryear{{Weaver}, {McCray}, {Castor}, {Shapiro} \&
  {Moore}}{{Weaver} et~al.}{1977}]{Weaver_et_1977}
{Weaver} R.,  {McCray} R.,  {Castor} J.,  {Shapiro} P.,    {Moore} R.,  1977,
  \apj, 218, 377

\bibitem[\protect\citeauthoryear{{Wesson}, {Barlow}, {Matsuura} \&
  {Ercolano}}{{Wesson} et~al.}{2015}]{Wesson_et_2015}
{Wesson} R.,  {Barlow} M.~J.,  {Matsuura} M.,    {Ercolano} B.,  2015, \mnras,
  446, 2089

\bibitem[\protect\citeauthoryear{{Wiener}, {Pfrommer} \& {Oh}}{{Wiener}
  et~al.}{2017}]{Wiener_et_2017}
{Wiener} J.,  {Pfrommer} C.,    {Oh} S.~P.,  2017, \mnras, 467, 906

\bibitem[\protect\citeauthoryear{{Xie}, {Yuan} \& {Ho}}{{Xie}
  et~al.}{2017}]{Xie_et_2017}
{Xie} F.-G.,  {Yuan} F.,    {Ho} L.~C.,  2017, \apj, 844, 42

\bibitem[\protect\citeauthoryear{{Yang}, {Gaspari} \& {Marlow}}{{Yang}
  et~al.}{2019}]{Yang_et_2019}
{Yang} H. Y.~K.,  {Gaspari} M.,    {Marlow} C.,  2019, \apj, 871, 6

\bibitem[\protect\citeauthoryear{{Yuan} \& {Narayan}}{{Yuan} \&
  {Narayan}}{2014}]{Yuan_Narayan_2014}
{Yuan} F.,  {Narayan} R.,  2014, \araa, 52, 529

\bibitem[\protect\citeauthoryear{{Zhang} \& {Davis}}{{Zhang} \&
  {Davis}}{2017}]{Zhang_Davis_2017}
{Zhang} D.,  {Davis} S.~W.,  2017, \apj, 839, 54

\end{thebibliography}

\label{lastpage}

\end{document}